\documentclass[aps,prb,twocolumn,superscriptaddress]{revtex4-2}

\usepackage{amsmath,amssymb,latexsym}
\usepackage{graphicx}
\usepackage{txfonts,mathrsfs}
\usepackage{comment}
\usepackage{xcolor}

\usepackage{float}

\begin{document}

\title{
Fermion quartets on the square lattice
}

\author{Dionys Baeriswyl}
\affiliation{Department of Physics, University of Fribourg, 1700 Fribourg, Switzerland}
\author{Francesco Petocchi}
\affiliation{Department of Physics, University of Fribourg, 1700 Fribourg, Switzerland}
\affiliation{Department of Quantum Matter Physics, University of Geneva, 1211 Geneva 4, Switzerland}
\author{Philipp Werner}
\affiliation{Department of Physics, University of Fribourg, 1700 Fribourg, Switzerland}

\begin{abstract}
We study a microscopic model for four spinless fermions on the square lattice which exhibits a quartet bound state in the strong coupling regime. The four-particle quantum states are analyzed using symmetry arguments and by introducing a zoo of relevant lattice animals. These considerations, as well as variational and exact diagonalization calculations demonstrate the existence of a narrow quartet band at small hopping and a first order transition to delocalized fermions at a critical hopping parameter, in qualitative contrast to, e.~g., the BCS-BEC crossover in the attractive Hubbard model. In the case of pure attraction, an intermediate phase is found, in which a more extended and presumably more mobile hybrid quartet dominates the ground state. We comment on the relevance of the spin degree of freedom and on the reasons why electron quartetting is rarely observed in real materials.  
\end{abstract}

\date{\today}


\maketitle

\section{Introduction}\label{sec:introduction}
Pairing, the central paradigm of superconductivity, has been amply confirmed experimentally. For example, the magnetic flux in superconductors is quantized in units of $\Phi_0=\frac{h}{2e}$ (and not $\frac{h}{e}$) and tunneling supercurrents are characterized by the Josephson constant $K_J=\frac{2e}{h}$. However, there exists no fundamental objection against building blocks consisting of $4, 6, 8,...$ electrons. In Yang's concept of off-diagonal long-range order the only requirement is that the number of fermions forming the ``basic group'' of particles be even \cite{Yang_62}.

Very recently, flux quantization in units of $\frac{h}{4e}$ (and even  $\frac{h}{6e}$) has been reported for the kagome lattice superconductor CsV$_3$Sb$_5$ \cite{Ge_22}. While below the critical temperature $T_c$ the conventional $\frac{h}{2e}$ period is observed in the magnetoresistance (Little-Parks effect), this oscillation vanishes above $T_c$, and new oscillations with a period $\frac{h}{4e}$ appear, taken as evidence for ``vestigial'' charge $4e$ superconductivity. Vestigial order can occur in systems with more than one potentially broken symmetry and is described by a composite order parameter, which may emerge if the primary order parameters vanish \cite{Fradkin_15,Fernandes_19, Agterberg_20}. An example of a system with more than one broken symmetry is a condensed state of Cooper pairs with finite momenta, a pair density wave, which is superconducting and also breaks translational symmetry. If both order parameters vanish, a condensation of Cooper quartets - products of Cooper pairs with opposite momenta - remains possible as a vestigial phase, leading to charge $4e$ superconductivity. Such a mechanism may be at work in the kagome material CsV$_3$Sb$_5$ \cite{Zhou_22}; the nature of the parent superconducting phase however remains to be clarified.

An interesting neutral four-fermion bound state is the positronium molecule, consisting of two electrons and two positrons. A long time ago, variational calculations have shown that this bound state is indeed stable \cite{Hylleraas_47, Kinghorn_93}, but experimental evidence for the positronium molecule has been reported only relatively recently \cite{Cassidy_07}. A closely related problem is that of two electrons and two holes in a semiconductor, which can form a bound state, a biexciton \cite{Moskalenko_00}. The most simple model of a semiconductor, with parabolic conduction and valence bands, is essentially the same as the Hamiltonian for electrons and positrons, except that the electron and hole masses are in general different (and may be anisotropic) and the Coulomb interaction is reduced by a dielectric constant. Variational calculations for the ground state of two electrons and two holes in fact have shown that the biexciton state is stable for arbitrary mass ratios \cite{Akimoto_72, Brinkman_73}. In a biexciton, attraction exists only between electrons and holes, and it is tempting to determine first the electron-hole bound state, the exciton. In a second step, the excitons are taken as bosons coupled by a residual attraction \cite{Nozieres_82}. Unfortunately, it is not easy to determine accurately the effective exciton-exciton interaction \cite{Combescot_08}.

A crucial question is whether biexcitons remain stable in a semiconductor with a finite density of electrons and holes. For Ge the preferred state at a large density has been shown to be an electron-hole plasma, both experimentally \cite{Thomas_73} and theoretically \cite{Combescot_72, Brinkman_73}, for early reviews see Refs.~\cite{Rice_78, Hensel_78}. Experimental evidence for biexcitons  has been reported for CuCl \cite{Grun_70}.  Luminescence spectra at high excitation density have been interpreted in terms of Bose-Einstein condensation of biexcitons \cite{Chase_79}, but this view has been challenged. A central point is that CuCl is a direct gap semiconductor. This means, on the one hand, that thermalization of electron-hole pairs is preceded by recombination, in contrast to indirect gap semiconductors such as Ge or Si, where the recombination time is much longer. On the other hand, the coupling to light can no longer be ignored and 
biexcitons become bipolaritons, which can be brought into a coherent state through resonant two-photon absorption with total momentum 0 \cite{Ivanov_98}. This produces a ``pump-induced'' out-of-equilibrium distribution of polaritons, which resembles Bose-Einstein condensation. 

More recently, the interest in excitons and biexcitons increased dramatically with the advent of coupled quantum well structures \cite{Butov_07}, where electrons and holes reside on two neighboring layers, which strongly enhances the exciton lifetime. A second boost occurred when atomically thin transition metal dichalcogenides became available \cite{Wang_18}. In monolayer WSe$_2$ the Coulomb interaction between electrons and holes is very strong and biexcitons have been created and detected \cite{You_15, Ye_18, Li_18}.

In nuclear physics, the $\alpha$-particle, a bound state of two protons and two neutrons, has played a central role, ever since its discovery at the end of the 19th century.
In contrast to the biexciton problem, where interactions are both simple and well defined (Coulomb),  
the nucleon-nucleon interactions responsible for the bound state of two protons and two neutrons are neither simple (they include tensor, spin-orbit and three-body forces) nor unique. In the early days of nuclear physics Yukawa's pion exchange \cite{Yukawa_35}  provided a mechanism for the strong intermediate-range attraction between nucleons. The pion model has become  very popular and still is used successfully. Experimental input (such as scattering data) is often required for fixing the parameters of the nucleon-nucleon potential.  Since the advent of Quantum Chromodynamics (QCD) a lot of effort has been invested to derive the nucleon-nucleon interaction {\it ab initio}, using chiral Effective Field Theory \cite{Machleidt_11, Hammer_20} or lattice QCD \cite{Briceno_18}. 

 Already in the Thirties, it has been proposed that nuclei consisting of $2n$ protons and $2n$ neutrons, $^8$Be, $^{12}$C, $^{16}$O, etc., consist of $\alpha$-particles \cite{Hafstad_38, Freer_18}. The nucleus $^{12}$C is then a triangle of $\alpha$-clusters, and $^{16}$O a tetrahedron. This molecular view point was not universally accepted because nuclei are compact objects, at least in the ground state. Hence, a mean-field approach was advocated, where the nucleons move independently in a spherically symmetric confining potential. This shell model was quite successful  \cite{Caurier_05}, especially for light nuclei with filled neutron or proton shells (magic nuclei), but unbiased studies suggest that in reality, shell and cluster structures coexist \cite{Itagaki_04}. For the enigmatic Hoyle state, an excited state (resonance) of $^{12}$C, which is believed to be responsible for the abundance of $^{12}$C in the universe \cite{Hoyle_54, Freer_14}, the naive shell model fails badly due to the large spatial extent of the state. The molecular point of view appears to be much more appropriate, although it is not firmly established how the three 
$\alpha$ clusters are correlated. Spatial clustering \cite{Freer_18}, a (Bose-Einstein) condensate
\cite{Schuck_17} and a superposition of quantum liquid and clustering \cite{Otsuka_22} have been proposed.
 
Quartets have also been discussed in the context of cold atoms, in particular for fermions with spin $\frac{3}{2}$, representing four hyperfine levels. The (solvable) one-dimensional model of $S=\frac{3}{2}$ fermions with attractive contact interaction, describing atoms confined to elongated traps, indeed exhibits quartetting \cite{Wu_05, Guan_09, Schlottmann_12}.  Its lattice version, the attractive Hubbard model (with on-site coupling $U<0$) has also been investigated and found to show quartetting \cite{Lecheminant_05, Lecheminant_08, Capponi_16}. This is easily understood in the large $|U|$ limit, where the system is well desribed by an ensemble of local quartets, similarly to the $S=\frac{1}{2}$ case, where we just have local pairs in this limit. 

In our study we have in mind electrons in layered materials, and therefore consider a square lattice and assume the on-site interaction to be repulsive. With nearest- and next-nearest-neighbor attraction and repulsion between third and fourth neighbors we favor quartets on the four sites of a plaquette. We limit ourselves mostly to four particles and determine both the ground state and the energy spectrum. For these investigations, the concept of quantum lattice animals turns out to be very useful in the limit of strong interactions. We first consider the spinless case, where we find a first order transition from a compact quartet at strong coupling to complete delocalization for weak coupling. Later we show that this picture does not change if spin is added. Therefore we do not find any Cooper quartets and there is no BCS-BEC crossover. Some issues of quartetting become particularly transparent in our model, such as the stability of real quartets, the problem of Bose-Einstein condensation and its competition with spatial ordering, or the relation between position and momentum space representations.  

The paper is organized as follows. Section~\ref{sec:model} defines the model, while Sec.~\ref{sec:states} introduces the lattice animals and provides a symmetry analysis. The weak-coupling regime is studied in Sec.~\ref{sec:weak} and the strong coupling regime in Sec.~\ref{sec:strong}. In Sec.~\ref{sec:transition} we show that the weak and strong coupling limits cannot be continuously connected, while Sec.~\ref{sec:intermediate} demonstrates the existence of an intermediate phase in the model with pure attraction. We briefly discuss the effect of spin in Sec.~\ref{sec:spin}, before presenting our conclusions in Sec.~\ref{sec:conclusions}.

\section{Model Hamiltonian and its classical limit}\label{sec:model}

\subsection{Hamiltonian}
We study spinless fermions described by the Hamiltonian   
\begin{align}\label{eq:ham}
H=H_0+H_{\mbox{\scriptsize int}}\, ,
\end{align}
where
\begin{align}
H_0=-t\sum_{\langle {\bf m,n}\rangle}\big(c_{\bf m}^\dag c_{\bf n}^{\phantom{}}+c_{\bf n}^\dag c_{\bf m}^{\phantom{}}\big)
\end{align}
represents the hopping of fermions between neighboring sites of a square lattice and $c_{\bf n}^\dag, c_{\bf n}^{\phantom{}}$ are the creation and annihilation operators for the spinless fermion at site ${\bf n}$. 
It is convenient to introduce the dimensionless hopping operator
\begin{align}\label{eq:hopping}
T=\sum_{\langle {\bf m,n}\rangle}\big(c_{\bf m}^\dag c_{\bf n}^{\phantom{}}+c_{\bf n}^\dag c_{\bf m}^{\phantom{}}\big)\, ,
\end{align}
such that $H_0=-tT$. The interaction term
\begin{align}\label{eq:hamiltonian}
H_{\mbox{\scriptsize int}}=&\,V_1\sum_{\langle {\bf m,n}\rangle}n_{\bf m}n_{\bf n}+V_2\sum_{\langle\! \langle {\bf m,n}\rangle\!\rangle}n_{\bf m}n_{\bf n}\nonumber\\
&+V_3\sum_{\langle\!\langle\!\langle {\bf m,n}\rangle\!\rangle\!\rangle}n_{\bf m}n_{\bf n}
+V_4\sum_{\langle\!\langle\!\langle\!\langle {\bf m,n}\rangle\!\rangle\!\rangle\!\rangle}n_{\bf m}n_{\bf n}\, ,
\end{align}
with $n_{\bf n}=c_{\bf n}^\dag c_{\bf n}^{\phantom{}}$ the density at site ${\bf n}$,  couples nearest ($V_1$), second ($V_2$), third ($V_3$) and fourth ($V_4$) neighbors.
In most of the following discussion, we consider a system with short-range attraction ($V_1<0,\, V_2<0$) and medium-range repulsion ($V_3> 0,\, V_4> 0$).
The square lattice has $L=L_x\times L_x$ sites and we use periodic boundary conditions. 

\begin{figure*}[ht]
\centering
\includegraphics[width=14cm,angle=0]{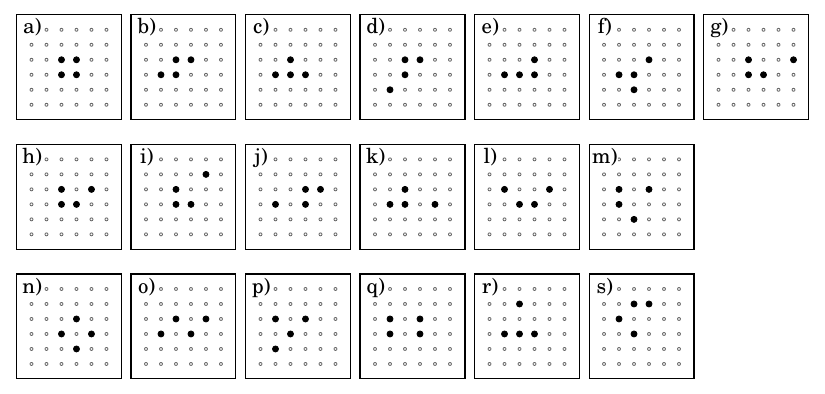}
\caption{Examples of four-particle configurations, arranged according to their energies, $E_a<E_b<...<E_s$,  for the coupling parameters (\ref{eq:parameters}).
}
\label{fig:animals}
\end{figure*}

\subsection{Origin of the attraction}
\label{sec:origin}

In a solid, short range attractive interactions can occur if the total charge on a plaquette couples to a local phonon, representing for example an apical oxygen mode. This situation is described by the plaquette Holstein Hamiltonian 
\begin{equation}
H_\text{plaquette}=\frac{\omega_0}{2}(X^2+P^2)+gNX+H_0, 
\end{equation}
where $X$ and $P$ represent the phonon displacement and momentum operators, $\omega_0$ is the phonon frequency, $g$ the electron-phonon coupling strength, $N=n_{{\bf m}_1}+n_{{\bf m}_2}+n_{{\bf m}_3}+n_{{\bf m}_4}$ the total charge on the plaquette, and $H_0$ the hopping term. We may decouple the electrons and phonons using a Lang-Firsov transformation \cite{Lang_62}: $\mathcal U=e^{-iPX_0}$ shifts the coordinate as $X\rightarrow X-X_0$, so that the transformed Hamiltonian reads $\tilde H_\text{plaquette}=\mathcal{U}^\dagger H\mathcal{U}=\frac{\omega_0}{2}((X-X_0)^2+P^2)-g(X-X_0)N+\mathcal{U}^\dagger H_0\mathcal{U}$. By choosing $X_0=\frac{g}{\omega_0}N$, one obtains a Hamiltonian without explicit electron-phonon coupling, but with an induced electron-electron interaction and a modified hopping term $\tilde H_0=\mathcal{U}^\dagger H_0\mathcal{U}$:
\begin{equation}
\tilde H_\text{plaquette}=\frac{\omega_0}{2}(X^2+P^2)-\frac{g}{2\omega_0}N^2+\tilde H_0.
\end{equation}  
Multiplying out the $N^2$ term yields, in the anti-adiabatic limit, an attractive interaction $\tilde U=-\frac{g^2}{\omega_0}$ between nearest-neighbor and next-nearest neighbor sites on the plaquette (plus a chemical potential shift). Assuming that each plaquette of a square lattice is coupled to such a phonon, we obtain an attractive nearest-neighbor interaction $V_1=2\tilde U$ and an attractive next-nearest neighbor interaction $V_2=\tilde U$.

\subsection{Classical limit}\label{sec:classical}
In the absence of hopping ($t=0$) the Hamiltonian (\ref{eq:ham}) is simply the energy $E$ for a set of occupation numbers $n_{\bf m}=0$ or 1. Figure~\ref{fig:animals} illustrates the four-particle configurations which will turn out to be relevant for small $t$. The energies of these ``lattice animals'' can be directly read off from the figure, namely $E=\sum_i \nu_i V_i$, where $\nu_1$ is the number of nearest-neighbor bonds and so on. Thus the energy of the ``plaquette state'' $a$ is 
$E_a=4V_1+2V_2$, while that of the  ``diamond'' $n$ is $E_n=4V_2+2V_3$.
In the absence of repulsive interactions all these lattice animals have negative energy for short-range attraction and are therefore stable with respect to configurations of well separated particles.
 However, they are unstable with respect to the addition of particles, leading to phase separation. Indeed, for a large droplet the energy per particle is roughly $2(V_1+V_2+V_3+2V_4)$, which is lower than that of any cluster in Fig.~\ref{fig:animals} for $V_1<0$, $V_2<0$ and $V_3=V_4=0$.
We will always assume $V_1$ to be the dominant attractive coupling, $V_1<V_2<0$. In this case the plaquette state $a$ has the lowest energy among all four-particle configurations for $V_3\ge0$, $V_4\ge 0$. Its stability with respect to aggregation requires repulsive interactions of a certain strength, namely $V_3+2V_4>-(\frac{1}{2}V_1+\frac{3}{4}V_2)$. 
We characterize the interaction by a single parameter $V>0$ and for most of the following analysis choose the couplings 
\begin{align}\label{eq:parameters}
V_1=-\frac{2}{3}V\, ,\quad V_2=-\frac{1}{3}V\, ,\quad V_3=\frac{1}{2}V\, ,\quad V_4=\frac{1}{4}V\, .
\end{align}
These ``canonical couplings" guarantee the stability of the plaquette state $a$ with respect to both deformation and aggregation. The ratio between $V_1$ and $V_2$ is furthermore consistent with the scenario discussed in Sec.~\ref{sec:origin}.

\begin{figure}[ht]
\centering
\includegraphics[width=6cm]{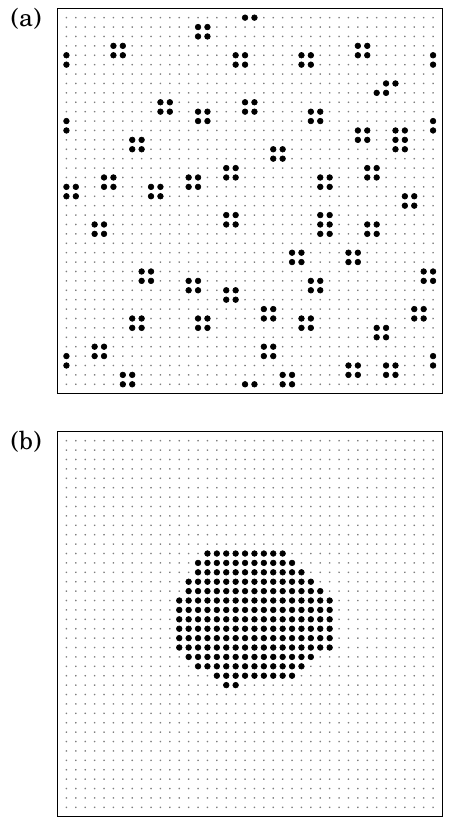}
\caption{Monte Carlo snapshots for 200 particles on $40\times 40$ sites at low temperature, $k_BT=0.1\, V$, and for periodic boundary conditions.  The parametrization (\ref{eq:parameters}) was used in panel (a), while panel (b) shows a snapshot for purely attractive couplings ($V_1=-\frac{2}{3}V\, ,V_2=-\frac{1}{3}V$,\, $V_3=V_4=0$). }
\label{fig:MC}
\end{figure}

Characteristic configurations for arbitrary density and nonzero temperatures can be readily obtained using Monte Carlo sampling. For a given number of particles, a Monte Carlo step consists in moving a randomly chosen particle from an occupied to an empty site, with a probability defined by the Metropolis algorithm. 
Two typical snapshots at low temperature are shown in Fig.~\ref{fig:MC}, one for the canonical parameters of Eq.~(\ref{eq:parameters}), the other for pure attraction. For canonical parameters the plaquettes clearly dominate, while phase separation occurs for pure attraction.

\section{Quantum states} \label{sec:states}
We now turn to the quantum problem of four spinless fermions on the square lattice. To describe the four-particle quantum states, we proceed in close analogy to the classification of single-particle wave functions in a periodic potential. Both point-group and translational symmetries will be used.

\subsection{Point-group symmetry} \label{subsec:symmetry}

The square lattice has the point-group symmetry $D_4$. The eight elements of this group correspond to the covering operations of a square, namely the identity ($e$), three rotations about the axis through the center perpendicular to the square, one by $\pi/2$ ($r_+$), one by $-\pi/2$ ($r_-$), one by $\pi$ ($r_\pi$), and four reflections, one with respect to the $x$ axis ($\sigma_x$), one with respect to the $y$ axis ($\sigma_y$), one with respect to the up-going diagonal ($\sigma_d$), and one with respect to the down-going diagonal ($\sigma'_d$). 

These transformations on the square lattice (and similarly in the Brillouin zone) induce unitary transformations in the Hilbert space through the rule
\begin{align}
c_{\bf n}\rightarrow c_{g{\bf n}}\, ,
\end{align}
where $g$ is any element of $D_4$. We will use the same notation for the transformations on the square lattice as for those in the Hilbert space.

The group has five conjugacy classes, namely $C_1=\{e\}$, $C_2=\{r_+,r_-\}$, $C_3=\{r_\pi\}$, $C_4=\{\sigma_x,\sigma_y\}$, $C_5=\{\sigma_d,\sigma'_d\}$. Therefore there are five inequivalent irreducible representations, four one-dimensional ($A_1,A_2,B_1,B_2$) and one two-dimensional ($E$). The character table is given in Tab.~\ref{tab:characters}.
\begin{table}[H]
\centering
$\begin{array}{l|rrrrr}
&C_1&C_2&C_3&C_4&C_5\\
\hline
A_1&1&1&1&1&1\\
A_2&1&1&1&-1&-1\\
B_1&1&-1&1&1&-1\\
B_2&1&-1&1&-1&1\\
E&2&0&-2&0&0\\
\end{array}$
\caption{Character table of the symmetry group $D_4$.}
\label{tab:characters}
\end{table}

A reducible representation $\Gamma$ with characters $ \chi(C_j)$ can be decomposed into irreducible representations using relations from group theory, namely
 \begin{align}
& \chi(C_j)=\sum_{k=1}^5a_k\, \chi^{(k)}(C_j)\, ,\nonumber\\
& 8a_k=\sum_{j=1}^5n_j\, \chi(C_j)\, \chi^{(k)*}(C_j)\, ,
 \end{align}
 where $n_j$ is the number of elements in the class $C_j$ and $k$ numbers the different irreducible representations. The decomposition is then specified by the formula
 \begin{align}\label{eq:decomposition}
 \Gamma=a_1A_1\oplus a_2A_2\oplus a_3B_1\oplus a_4B_2\oplus a_5E\, .
 \end{align}
 
The fixed point of the point-group operations on the square lattice can be a site or the center of a plaquette. 

\subsection{Quantum lattice animals}
The states
\begin{align}\label{eq:states_position}
\vert {\bf n}_1,{\bf n}_2,{\bf n}_3,{\bf n}_4\rangle:=c_{{\bf n}_1}^\dag c_{{\bf n}_2}^\dag c_{{\bf n}_3}^\dag c_{{\bf n}_4}^\dag\vert 0\rangle\, ,
\end{align}
with distinct sites ${\bf n}_1,...,{\bf n}_4$, yield a natural basis for fermion quartets on the square lattice. However, if only relatively compact configurations are relevant, such as those of Fig. \ref{fig:animals}, other labels than the locations of the individual particles are more useful, for instance the species $S$ ($S=a,b,c,...$ for the lattice animals of Fig.~\ref{fig:animals}), the center of mass, and the orientation of the quartets.

As a first example, we consider the plaquette state (the structure $a$ in Fig. \ref{fig:animals})
\begin{align}\label{eq:plaquette}
\vert a{\bf n}\rangle:=\vert {\bf n},{\bf n+i},{\bf n+i+j},{\bf n+j}\rangle\, ,
\end{align}
where ${\bf i}$ and ${\bf j}$ are unit vectors in the $x$- and $y$-direction, respectively. The natural fixed point of the point-group operations is the center of the plaquette. In this case the state  $\vert a{\bf n}\rangle$ remains invariant under the application of $e$, $r_\pi$, $\sigma_x$, $\sigma_y$ and
changes sign for $r_\pm$, $\sigma_d$, $\sigma_d'$. Therefore $\vert a{\bf n}\rangle$ transforms according to the irreducible representation $B_1$ (see Tab.~\ref{tab:characters}).

We now consider a generic state, i.e., a species $S$ with a well-defined location ${\bf n}$ and without any particular symmetry (such as $S=d, e, g$ in Fig. \ref{fig:animals}). In this case there are 8 different states 
$\vert S\ell {\bf n}\rangle$, $\ell=1,...,8$, linked to each other by the point-group transformations. There are $L$ different locations ${\bf n}$ and 8 different orientations $\ell$, therefore the states $\vert S\ell{\bf n}\rangle$ define an $8L$-dimensional subspace for a fixed species $S$.
As shown in Appendix \ref{app:irreps}, the action of group operations on some arbitrarily chosen initial state (\ref{eq:states_position}) yields a reducible 8-dimensional representation $\Gamma$ with decomposition
\begin{align}\label{eq:decomposition1}
\Gamma=A_1\oplus A_2\oplus B_1 \oplus B_2\oplus 2E\, .
\end{align}
Often it is advantageous to label the states according to the irreducible representations $\Gamma$, $\vert S\ell {\bf n}\rangle\rightarrow\vert S\Gamma{\bf n}\rangle$ (transformation (\ref{eq:irreps})).

The list of Fig.~\ref{fig:animals} contains also symmetric examples. For instance, the structures $c$ and $l$ are symmetric with respect to a reflection about the $y$-axis, while $f$ and $i$ are symmetric with respect to a reflection about the up-going diagonal. In these cases, there are only four different states related to each other by rotations and reflections. These states generate four-dimensional representations, which are decomposed into irreducible representations as shown in Tab.~\ref{tab:irreps}. The table also contains the two fully symmetric clusters $a$ and $n$, which generate one-dimensional representations. 
\begin{table}[H]
\centering
$
\begin{array}{cc}
\mbox{Species}&\mbox{Decomposition}\\
\hline
a&B_1\\
c&A_2\oplus B_2\oplus E\\
f&A_2\oplus B_1\oplus E\\
l&A_1\oplus B_1\oplus E\\
n&B_2
\end{array}
$
\caption{Irreducible representations of some symmetric quartets. The species are labeled as in Fig. \ref{fig:animals}. The point-group operations have been defined according to the natural quartet centers.
}
\label{tab:irreps}
\end{table}

Structures without a well-defined center are special. As an example we consider the case $b$ of Fig. \ref{fig:animals}. We can arbitrarily put such a species somewhere on the lattice, take one of the internal sites as its center and generate seven other states by applying rotations and reflections with respect to this center.
Naively, we would expect to obtain $8L$ independent states in this way. However, it is easy to see that there are only $4L$ different states of this species, by noticing that there are two linearly independent states for each nearest-neighbor bond. With $2L$ such bonds we arrive at $4L$ states. The point is that states produced by the application of symmetry operations on neighboring sites can be equal. Still considering the lattice animal $b$ in Fig.~\ref{fig:animals}, one notices that a $\pi$-rotation around one of the middle sites reproduces the orientation but not the location, which is shifted by one lattice constant. Thus this species has a symmetry involving both a rotation and a translation, as in non-symmorphic crystals.
In the following, structures with a well defined location (at a site or at the center of a plaquette) will be called {\it symmorphic}, while those with an ambiguous location will be named {\it non-symmorphic}. If we include all sites and all orientations of a non-symmorphic species we obtain an over-complete set of states. There are two ways of solving this problem. Either one could choose only the sites of one sublattice, say sublattice $\cal{A}$ for which $n_x+n_y$ is even, as centers, or reduce the number of orientations at each site by a factor of 2. We will use mostly the second procedure, which is detailed in Appendix \ref{app:irreps}.

\subsection{Bloch states of lattice animals}
The states $\vert S\ell {\bf n}\rangle$ introduced above are orthogonal for symmorphic structures,
\begin{align}
\langle S\ell{\bf n}\vert S'\ell'{\bf n'}\rangle=\delta_{S,S'}\delta_{\ell,\ell'}\delta_{\bf n,n'}
\end{align}
and thus are the analogue of Wannier states for a particle in a periodic potential. The corresponding Bloch states
\begin{align}\label{eq:states_Bloch}
\vert S\ell {\bf k}\rangle=\frac{1}{\sqrt{L}}\sum_{\bf n}e^{-i{\bf k\cdot n}}\vert S\ell {\bf n}\rangle\, ,
\end{align}
with ${\bf k}$-vectors in the Brillouin zone, are also orthogonal,
\begin{align}
\langle S\ell{\bf k}\vert S'\ell'{\bf k'}\rangle=\delta_{S,S'}\delta_{\ell,\ell'}\delta_{\bf k,k'}\, .
\end{align}

For non-symmorphic structures, the situation is similar to that of mole\-cular-orbital theory in quantum chemistry, where one faces the problem of non-orthogonality 
of atomic wave functions for neighboring atoms \cite{Pople_67}. Here, we simply reduce the over-complete basis, as mentioned above and worked out in Appendix \ref{app:irreps}.
For instance, in the case of species $b$ we limit ourselves to four orientations $\ell$, or to four irreducible representations $\Gamma$ if these are used as labels.

\begin{figure*}
\centering
\includegraphics[width=10cm]{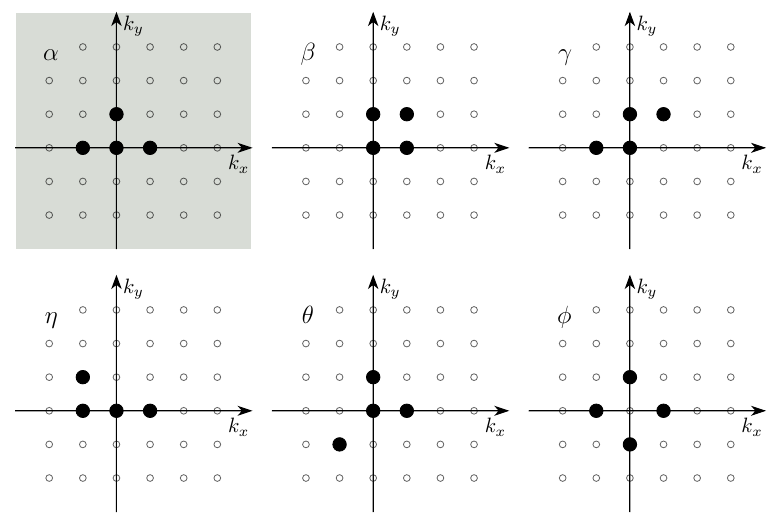}
\caption{Ground state of $H_0$ ($\alpha$) and first excited states ($\beta$ to $\phi$) in ${\bf k}$-space.}
\label{fig:levels}
\end{figure*}

\subsection{Momentum space}\label{subsec:momentum}

The states described above are useful if only a small ``zoo'' of lattice animals is required, i.e., if the interaction term $H_{\mbox{\scriptsize int}}$ dominates and only clusters with low classical energies have to be retained. In the opposite limit with dominant kinetic energy $H_0$, the important states are the lowest eigenstates of $H_0$. In this case it is natural to use the momentum representation,
\begin{align}
c_{\bf n}=\frac{1}{\sqrt{L}}\sum_{\bf k}e^{i{\bf k\cdot n}}c_{\bf k}
\end{align}
together with periodic boundary conditions, i.e., 
\begin{align}
{\bf k}=(\mu p,\nu p)\, ,\,\quad p=\frac{2\pi}{L_x}\, ,\quad -\frac{L_x}{2}<\mu,\nu\le\frac{L_x}{2}\, .
\end{align}
In this representation the hopping term is diagonal,
\begin{align}
H_0=\sum_{\bf k}\varepsilon_{\bf k}c_{\bf k}^\dag c_{\bf k}^{\phantom{}}\, ,
\end{align}
with the tight-binding spectrum
\begin{align}
\varepsilon_{\bf k}=-2t(\cos k_x+\cos k_y)\, ,
\end{align}
while the interaction is given by
\begin{align}
H_{\mbox{\scriptsize int}}=\frac{1}{2L}\sum_{{\bf k}_1,{\bf k}_2,{\bf q}} V({\bf q})\, 
c_{{\bf k}_1}^\dag  c_{{\bf k}_2}^\dag c_{{\bf k}_2+{\bf q}}^{\phantom{}}c_{{\bf k}_1-{\bf q}}^{\phantom{}}\, ,
\end{align}
with the momentum-dependent coupling
\begin{align}\label{eq:potential}
&V({\bf q})=\sum_{\bf n}e^{-i{\bf q\cdot n}}V({\bf n})\nonumber\\
&=2V_1(\cos q_x+\cos q_y)+4V_2\cos q_x\cos q_y\nonumber\\
&+2V_3(\cos 2q_x+\cos 2q_y)+4V_4(\cos q_x \cos 2q_y+\cos 2q_x\cos q_y)\, .
\end{align}

The point-group operations in position space induce corresponding transformations in momentum space. For a center at the site $(0,0)$ we find
\begin{align}
gc_{\bf k}=\frac{1}{\sqrt{L}}\sum_{\bf n'}e^{-i(g{\bf k})\cdot {\bf n'}}c_{\bf n'}=c_{g{\bf k}}\, ,
\end{align}
where the invariance of the scalar product has been used, \mbox{${\bf k}\cdot (g^{-1}{\bf n'})=(g{\bf k})\cdot {\bf n'}$}. (For other centers, additional phase factors are produced.)

The four-particle eigenstates of $H_0$ are
\begin{align}\label{eq:states_momentum}
\vert{\bf k}_1,{\bf k}_2,{\bf k}_3,{\bf k}_4\rangle:=c_{{\bf k}_1}^\dag c_{{\bf k}_2}^\dag c_{{\bf k}_3}^\dag c_{{\bf k}_4}^\dag\vert0\rangle\, ,
\end{align}
where all wave vectors are distinct. Those with the lowest eigenvalue have wave vectors 
${\bf k}_1={\bf 0}$ and ${\bf k}_2,{\bf k}_3,{\bf k}_4$ chosen among $(p,0),(-p,0),(0,p)$ and $(0,-p)$. This can be done in four different ways, for instance as in the structure $\alpha$ of Fig.~\ref{fig:levels}. The application of $\pi/2$ rotations yields the three other orientations. One obtains a four-dimensional representation $\Gamma$, exactly as for the species $c$ of 
Fig. \ref{fig:animals}, with the same decomposition as in Table \ref{tab:irreps}, namely
\begin{align}
 \Gamma=A_2\oplus B_2\oplus E\, .
\end{align}

The state (\ref{eq:states_momentum}) is an eigenstate of the crystal momentum
\begin{align}
{\bf P} :=\sum_{\bf k}{\bf k}c_{\bf k}^\dag c_{\bf k}^{\phantom{}}
\end{align}
with eigenvalue ${\bf k}_1+{\bf k}_2+{\bf k}_3+{\bf k}_4$. It is worthwhile to mention that the Bloch states for the lattice animals (\ref{eq:states_Bloch}) are not eigenstates of ${\bf P}$ with eigenvalue ${\bf k}$, but superpositions of momentum eigenstates with eigenvalues ${\bf k+K}$, where ${\bf K}$ are reciprocal lattice vectors.

\section{Weak coupling}\label{sec:weak}
For weak interactions, $V\ll t$, the momentum eigenstates (\ref{eq:states_momentum}) are expected to provide a useful starting point for calculating the low-energy states of the Hamiltonian. In a first step, we limit ourselves to the reduced subspace defined by Fig. \ref{fig:levels} and perform analytical calculations. In a second step, we present results from Exact Diagonalization for an $8\times 8$ lattice. Both methods yield a four-fold degenerate ground state, in agreement with the symmetry arguments.

\subsection{Hartree-Fock}\label{subsec:spectrum}
To calculate the energy spectrum using perturbation theory (or variational methods) for weak coupling we need the matrix elements
\begin{align}\label{eq:matrix}
\langle{\bf k}'_1,{\bf k}'_2,{\bf k}'_3,{\bf k}'_4\vert H_{\mbox{\scriptsize int}}\vert{\bf k}_1, {\bf k}_2,{\bf k}_3,{\bf k}_4\rangle\, .
\end{align}
It is straightforward to establish some simple rules, using both translational and $D_4$ symmetries together with Wick's theorem:
\begin{enumerate}
\item The matrix element (\ref{eq:matrix}) vanishes if $\sum_i {\bf k}_i\neq \sum_i {\bf k}'_i+{\bf K}$, where ${\bf K}$ is any reciprocal lattice vector, or if the two states belong to two different irreducible representations of $D_4$.
\item Diagonal matrix elements are equal if the states are related by a symmetry transformation $g\in D_4$, i.e.,
\begin{align}
\langle\Psi\vert H_{\mbox{\scriptsize int}}\vert\Psi\rangle=\langle g\Psi\vert H_{\mbox{\scriptsize int}}\vert g\Psi\rangle\, .
\end{align}
\item Off-diagonal matrix elements $\langle\Psi\vert H_{\mbox{\scriptsize int}}\vert\Psi'\rangle$ vanish except if $\vert\Psi\rangle$ and $\vert\Psi'\rangle$ 
have two or four common wave vectors.
\end{enumerate}

We now apply these rules to the eigenstates of $H_0$ with the two lowest eigenvalues. These states are illustrated in Fig. \ref{fig:levels}. The lowest level at
\begin{align}\label{eq:level1}
E_0=-2t(5+3\cos p)
\end{align}
(where $p=\frac{2\pi}{L_x}$) is four-fold degenerate (species $\alpha$), while the next higher eigenvalue
\begin{align}\label{eq:level2}
E_1=-8t(1+\cos p)
\end{align}
has 25 eigenstates (species $\beta$ to $\phi$). The representations generated by these states are detailed in Table \ref{tab:states}. The total crystal momenta ${\bf P}$ of the different species are all different except for those labeled by $\theta$ and $\phi$, for which ${\bf P}$ vanishes. But the decompositions of the representations generated in these two cases do not contain any common element. Therefore, according to rule 1, all the matrix elements between states belonging to different species vanish and we can consider each one independently to determine the first-order corrections to the energies.  The corrections are proportional to $\frac{1}{L}$, and thus vanish in the thermodynamic limit. This is of course expected because in this limit we have a (four-particle) scattering problem, where the energy is asymptotically that of independent particles. Explicit expressions are presented in Appendix \ref{app:perturbation}.

Because our procedure is similar to the Hartree-Fock approximation, we will use the label HF for these results.

\begin{table}[t]
\centering
$\begin{array}{cccc}
{\mbox{Species}}&\vert {\bf P}\vert&\mbox{States}&{\mbox{Decomposition}}\\
\hline
\alpha&p&4&A_2\oplus B_2\oplus E\\
\beta&\sqrt{8}p&4&A_2\oplus B_1\oplus E\\
\gamma&2p&8&A_1\oplus A_2\oplus B_1\oplus B_2\oplus 2E\\
\eta&\sqrt{2}p&8&A_1\oplus A_2\oplus B_1\oplus B_2\oplus 2E\\
\theta&0&4&A_2\oplus B_1\oplus E\\
\phi&0&1&B_2
\end{array}$
\caption{Crystal momenta, number of states and irreducible representations generated by the low-lying four-particle eigenstates of $H_0$. The labels $\alpha$ to $\phi$ refer to the different structures shown in 
Fig.~\ref{fig:levels}.}
\label{tab:states}
\end{table}

\subsection{Exact Diagonalization}
To determine the energy spectrum for arbitrary coupling strength, we use exact diagonalization (ED). The Hamiltonian (\ref{eq:ham})
has been implemented for four particles on a $8\times 8$ square lattice, using periodic boundary conditions. The 
Hilbert space of all possible configurations $\left|\phi_i\right\rangle $
has dimension $N_{H}=635376$. The corresponding matrix
representation is brought to a tridiagonal form using
the Lanczos algorithm \cite{Lin_93} and diagonalized with state-of-the-art
parallel routines providing a subset of eigenvectors $|\Psi^{(n)}\rangle $
up to a user-defined tolerance $(10^{-15})$. Each eigenvector
is explicitly stored as a superposition over the basis states via the complex coefficients $\alpha_{i}^{\left(n\right)}$,
\begin{equation}
|\Psi^{\left(n\right)}\rangle =\sum_{i=1}^{N_H}\alpha_{i}^{\left(n\right)}\left|\phi_i\right\rangle \, ,
\end{equation}
which allows us to evaluate arbitrary operators ${\mathcal{O}}={c}_{\mathbf{n}}^{\dagger}{c}_{\mathbf{m}}^{\dagger}\ldots{c}_{\mathbf{m'}}^{\phantom{}}{c}_{\mathbf{n'}}^{\phantom{}}$
that conserve the number of particles. 

\begin{figure}[t]
\centering
\includegraphics[width=\columnwidth,angle=0]{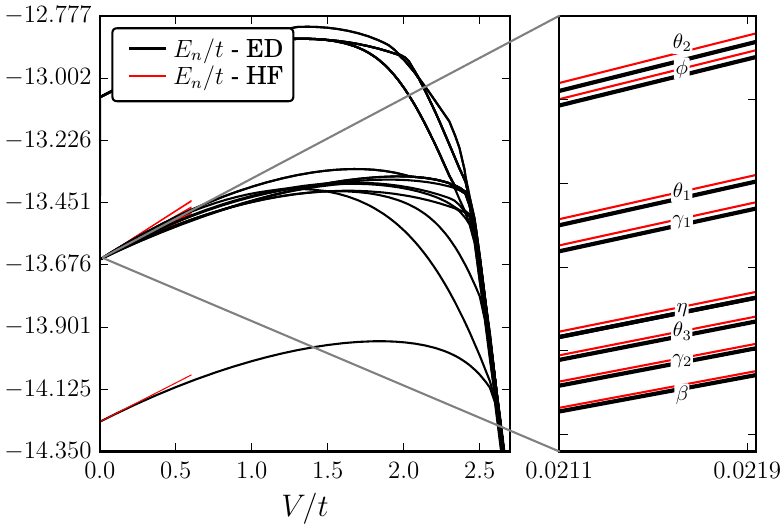}
\caption{Low-energy eigenvalues as functions of the coupling strength $V$. The values of the interaction parameters $V_1,..., V_4$ have been chosen according to Eq. (\ref{eq:parameters}). ED results are shown in the left panel and compared to the HF results (red lines) in the right panel.}
\label{fig:spectrum}
\end{figure}

Figure~\ref{fig:spectrum} shows the ED results and compares them with perturbation theory (Hartree-Fock). In this way we can attribute the low-lying energy levels to the species of Fig. \ref{fig:levels}. We notice that the evolution of the spectrum approximately follows the predictions of perturbation theory for $V<t$. However, the behavior changes markedly for $V>2t$, where several levels merge into a narrow band, which eventually becomes the dominant low-energy feature. 

For $V \lesssim t$ the ground state $\vert\Psi_0\rangle$ is dominated by the $\alpha$ species and the states generated by applying $\frac{\pi}{2}$ rotations are linearly independent. This, together with the symmetry $Hg\vert\Psi_0\rangle=gH\vert\Psi_0\rangle$, is the reason why the ground state is fourfold degenerate.

\section{Strong coupling}\label{sec:strong}
We now turn to the opposite limit of dominant interactions. We expect to find a ground state which resembles the classical quartet of Section \ref{sec:classical}. Quantum fluctuations have two effects. On the one hand, they lead to a spreading of the wave function, on the other hand, they restore the translational invariance by delocalizing the quartet as a whole. We use both a variational approach, where these two effects are treated in two steps, and exact diagonalization, where the bound quartet state 
corresponds to 
a narrow band well separated from the rest of the energy spectrum. The comparison of the two methods shows that the variational wave function represents very well the ground state in the region where the four particles are bound. 

\subsection{Variational ansatz}\label{subsec:variational}
We use a variational ansatz which has been introduced in the framework of the large-$U$ Hubbard model \cite{Baeriswyl_87} and applied to a Hamiltonian for spinless fermions with long-range Coulomb interaction to describe the melting of the ``generalized Wigner crystal'' \cite{Valenzuela_03}. In the present context, the ansatz is defined as
\begin{align}\label{eq:ansatz}
\vert\Psi{\bf n}\rangle=e^{\kappa T}\vert a{\bf n}\rangle\, ,
\end{align}
where $\vert a{\bf n}\rangle$ is the plaquette state (\ref{eq:plaquette}), $T$ is the hopping operator (\ref{eq:hopping}), and the variational parameter $\kappa$ controls the spreading of the wave function. The evaluation of   the expectation value
\begin{align}\label{eq:expectation}
E(\kappa)=\frac{\langle\Psi{\bf n}\vert H\vert\Psi{\bf n}\rangle}{\langle\Psi{\bf n}\vert\Psi{\bf n}\rangle}
\end{align}
is carried out explicitly in Appendix \ref{app:variational}. 
The expectation value of the hopping operator is given by the simple formula
\begin{align}\label{eq:kinetic}
\frac{\langle\Psi{\bf n}\vert T\vert\Psi{\bf n}\rangle}{\langle\Psi{\bf n}\vert\Psi{\bf n}\rangle}=-16\tau_1\, \frac{\tau_0-\tau_2}{\tau_0^2-\tau_1^2}\, ,
\end{align}
where
\begin{align}
\tau_\nu:=&\frac{1}{L_x}\sum_{k_x}e^{4\kappa\cos k_x}\cos^\nu k_x,\, \nu=0,1,2.
\end{align}
The expectation value of the interaction, Eq. (\ref{eq:energy1}), is a (complicated) function 
of {\it one-dimensional} amplitudes
\begin{align}
f_{n_\alpha}=\frac{1}{L_\alpha}\sum_{k_\alpha}e^{2\kappa \cos k_\alpha}\cos k_{\alpha}n_{\alpha}\, ,\quad \alpha=x,y.
\end{align}
This dimensional reduction facilitates greatly the computations, especially in the thermodynamic limit, where $f_n$ decreases exponentially with $\vert n\vert$ for $\kappa>0$, so that only a limited number of these amplitudes is required. For small lattices, it is advantageous to use the representation 
in momentum space, Eq. (\ref{eq:energy2}).

\begin{figure}[t]
\centering
\includegraphics[width=\columnwidth]{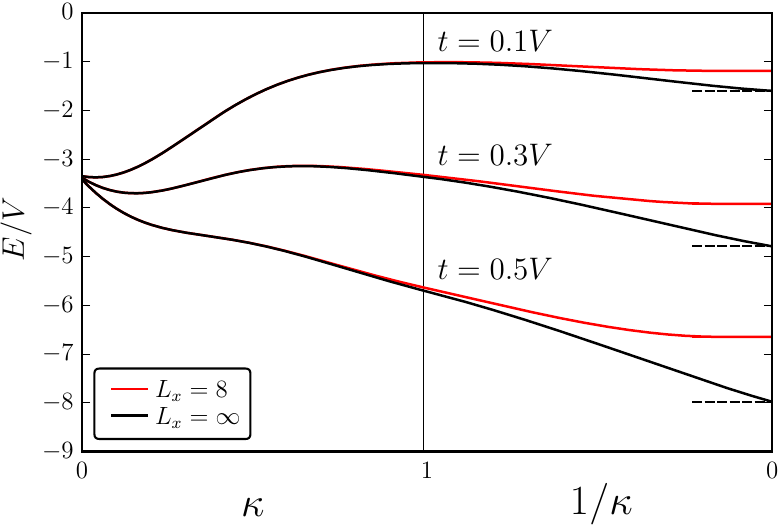}
\caption{Variational energy as a function of $\kappa$ for the coupling parameters (\ref{eq:parameters}) with $\frac{t}{V}=0.1$, $0.3$ and $0.5$ (top to bottom). Black lines show data for the thermodynamic limit and red lines are results obtained for $L_x=8$. The dashed lines represent the corresponding ground-state energies of $H_0$.}
\label{fig:variational}
\end{figure}

Figure~\ref{fig:variational} shows $E(\kappa)$ for different values of $t$. For $t=0.1\, V$ a global minimum occurs at a very small value of $\kappa$. A minimum is again found for 
$t=0.3\, V$, but it is only local and the global minimum has jumped to $\kappa=\infty$. For $t=0.5\, V$ no minimum is obtained for finite values of $\kappa$. The curves for $L_x=8$ and $L_x=\infty$ are almost on top of each other for $\kappa<1$, but differ appreciably for $\kappa\gg1$. For $\kappa\rightarrow\infty$ and $L_x=\infty$, the energy tends to $E_0=-16t$, the lowest eigenvalue of $H_0$. However, as shown in Appendix \ref{app:variational}, for a finite lattice the state reached asymptotically, the ``fully projected'' state, is a superposition of the structures $\beta,\gamma,\eta$ of Fig.~\ref{fig:levels}, of $B_1$ symmetry and with energy $E_1>E_0$.

\begin{figure}
\centering
\includegraphics[width=\columnwidth]{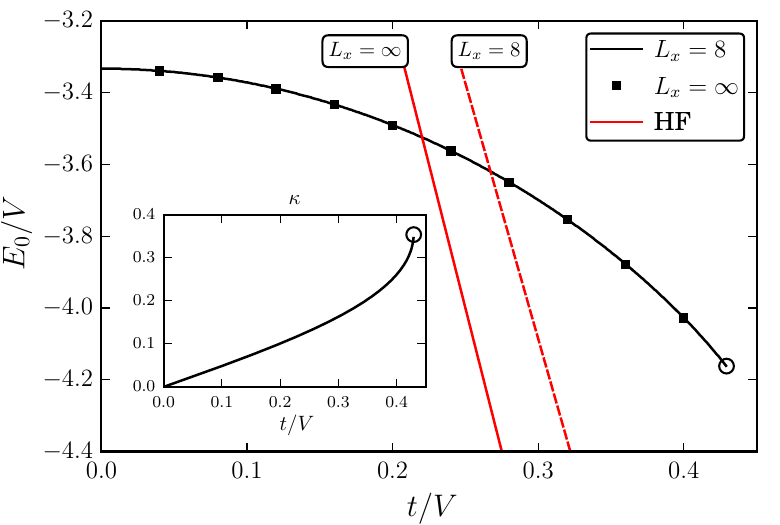}
\caption{Variational ground-state energies as functions of $t/V$ for the coupling parameters (\ref{eq:parameters}). Results for the ansatz (\ref{eq:ansatz}) are shown in black, with the optimized values of $\kappa$ given in the inset. Circles mark the critical value of $t$, beyond which no local minimum is found.
Red lines represent the Hartree-Fock results of Appendix \ref{app:perturbation}.}
\label{fig:emin}
\end{figure}

The minimization of $E(\kappa)$ yields the variational ground-state energy shown in Fig.~\ref{fig:emin}. For very small values of $t$, the bound quartet state has lower energy than the scattering state, while the latter is stabilized above a critical value $t_c\approx 0.220\, V$, where a first-order transition due to level crossing occurs. 

The role of $\kappa$ in the spreading of the wave function is neatly illustrated by the density, which is given by the simple expression (for details see Appendix \ref{app:density})

\begin{align}\label{eq:density}
\frac{\langle\Psi\vert c_{\bf n}^\dag c_{\bf n}^{\phantom{}}\vert\Psi\rangle}{\langle\Psi\vert\Psi\rangle}
=\frac{1}{(\tau_0^2-\tau_1^2)^2}\big[\tau_0(f^2_{n_x}+f^2_{n_x-1})-2\tau_1f_{n_x}f_{n_x-1}\big]\,\, \nonumber\\
\qquad \times \big[\tau_0(f^2_{n_y}+f^2_{n_y-1})-2\tau_1f_{n_y}f_{n_y-1}\big]\, .
\end{align}

\subsection{Linear spreading}\label{sec:linearspread}

 For hopping parameters which do not destabilize the bound quartet, the variational parameter $\kappa$ is very small, $\kappa\lesssim 0.11$. Therefore we can use a simplified ansatz
by expanding Eq.~(\ref{eq:ansatz}) to first order in $\kappa$,
\begin{align}\label{eq:linear}
\vert\Psi{\bf n}\rangle=(1+\kappa T)\vert a{\bf n}\rangle\, .
\end{align}
 A useful concept is that of sublattice parity, which we define as
\begin{align}
P_s:=e^{\frac{i\pi}{2}\sum_{\bf n}(-1)^{n_x+n_y}c_{\bf n}^\dag c_{\bf n}^{\phantom{}}}\, .
\end{align}
The square lattice consists of two sublattices, one with even, the other with odd values of $n_x+n_y$. For states $\vert{\bf n}_1,{\bf n}_2,{\bf n}_3,{\bf n}_4\rangle$, 
$P_s=+1$ if an even number of occupied sites ${\bf n}_1,...,{\bf n}_4$ belongs to either sublattice, otherwise $P_s=-1$.
Therefore $P_s=1$ for clusters $a$, $b$, $e$, $f$, $g$, \ldots, and $P_s=-1$ for clusters $c$, $d$, $h$, \dots \, of Fig.~\ref{fig:animals}.   
The application of the operator $T$ changes the sublattice parity, hence
$\langle a{\bf n}\vert T^k\vert a{\bf n}\rangle$ vanishes if $k$ is odd, and we get
\begin{align}
&\langle\Psi{\bf n}\vert\Psi{\bf n}\rangle=1+\kappa^2\langle a{\bf n}\vert T^2\vert a{\bf n}\rangle, 
\quad \langle\Psi{\bf n}\vert T\vert\Psi{\bf n}\rangle=2\kappa\langle a{\bf n}\vert T^2\vert a{\bf n}\rangle\, .
\end{align}

It is obvious from Fig.~\ref{fig:animals} that the application of $T$ to the plaquette state yields states of the species $h$, and indeed one finds
\begin{align}
T\vert a{\bf n}\rangle=\sqrt{8}\vert hB_1{\bf n}\rangle\, ,
\end{align}
where we have used the notation of Eq. (\ref{eq:irreps}), and therefore
\begin{align}
&\langle\Psi{\bf n}\vert\Psi{\bf n}\rangle=1+8\kappa^2, 
\quad \langle\Psi{\bf n}\vert T\vert\Psi{\bf n}\rangle=16\kappa\, .
\end{align}
The states $\vert a{\bf n}\rangle,\vert hB_1{\bf n}\rangle$ are eigenstates of $H_{\mbox{\scriptsize int}}$ with eigenvalues
\begin{align}
E_a&=4V_1+2V_2, \\
E_h&=2V_1+2V_2+V_3+V_4\, .
\end{align}
Collecting the various terms, we arrive at the variational energy
\begin{align}
E(\kappa)=E_h+\frac{E_a-E_h-16\kappa t}{1+8\kappa^2}\, .
\end{align}
Its extrema are at 
\begin{align}\label{eq:kappa}
\kappa=\frac{1}{16t}\left(E_a-E_h\pm\sqrt{(E_a-E_h)^2+32\, t^2}\right)
\end{align}
with values
\begin{align}\label{eq:eigenvalues}
E=\frac{1}{2}\left(E_a+E_h\mp\sqrt{(E_a-E_h)^2+32 t^2}\right)\, .
\end{align}

For the ground state and a small enough hopping parameter, $t\ll E_h-E_a$, Eqs. (\ref{eq:kappa}) and (\ref{eq:eigenvalues}) simplify to
\begin{align}
\kappa\approx\frac{t}{E_h-E_a}\, ,\qquad E\approx E_a-\frac{8t^2}{E_h-E_a}\, .
\end{align}
This is the amplitude for spreading the four-particle wave function to neighboring sites of the plaquette.

An alternative route to Eq. (\ref{eq:eigenvalues}) would be to diagonalize $H$ in the subspace spanned by the two states  $\vert a{\bf n}\rangle,\vert hB_1{\bf n}\rangle$. In fact, this yields a 
$2\times 2$ matrix with eigenvalues as in Eq. (\ref{eq:eigenvalues}).

\subsection{Quartet band}\label{sec:zoo}

So far, we have considered spatially localized four-particle states, which break the translational symmetry. To restore this symmetry, we now turn to the Bloch states (\ref{eq:states_Bloch}). All matrix elements $\langle S\ell{\bf k}\vert T\vert S\ell'{\bf k'}\rangle$ vanish (because each species $S$ has a definite sublattice parity, which is changed by the hopping operator). Therefore we need superpositions of different species to obtain a nonzero kinetic energy. The linear ansatz (\ref{eq:linear}) includes species $a$ and $h$, but still the matrix element of $T$ between two such states located on different sites ${\bf n\neq n'}$ vanishes. In fact, we need four hopping operators to move a plaquette state from a site ${\bf n}$ to a neighboring site ${\bf n'}$. 

The lattice animals of Fig.~\ref{fig:animals} have been generated by applying $T$ and $T^2$ to the plaquette state and $T$ to both $\vert b\ell{\bf n}\rangle$ and $\vert c\ell{\bf n}\rangle$. Choosing all these states as a basis, we obtain a $100\times 100$ Bloch matrix. Its diagonalization yields a rich band structure, of which only the low-energy part is expected to yield a good approximation for small $t$. At higher energies lattice animals not included in our basis become relevant, such as separated pairs or singly ionized quartets.

\begin{figure}[t]
\centering
\includegraphics[width=\columnwidth]{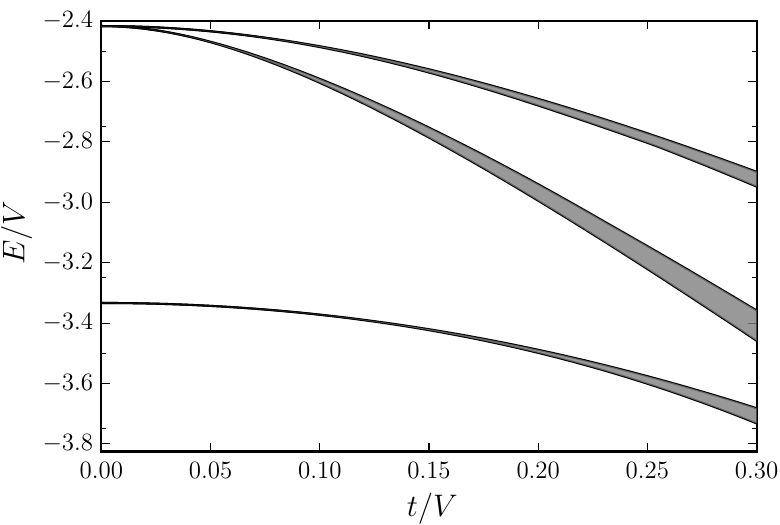}
\caption{Energy spectrum for the restricted set of lattice animals and canonical coupling parameters~\eqref{eq:parameters}.}
\label{fig:bandstructure_canonical}
\end{figure}

\begin{figure}[ht]
\centering
\includegraphics[width=\columnwidth]{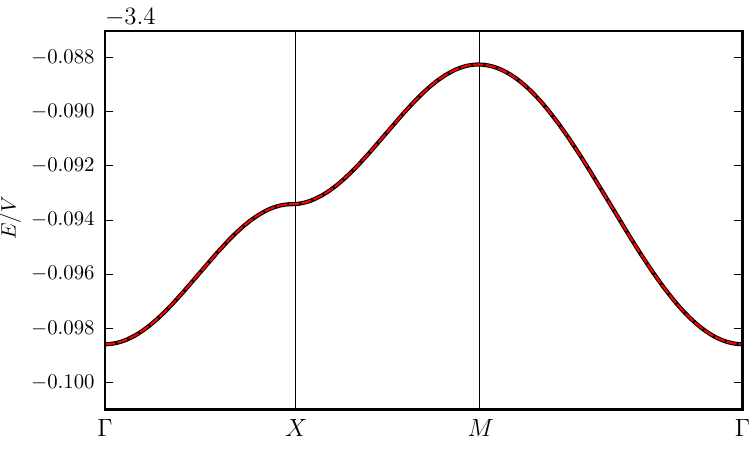}
\caption{Lowest eigenvalue of the Bloch matrix (solid line) for $t=0.2\, V$ and canonical coupling parameters, Eq. (\ref{eq:parameters}), along specific lines in the Brillouin zone. The dashed line is a tight-binding fit with an effective nearest-neighbor hopping parameter $t_{\mbox{\scriptsize eff}}\approx 0.00129\, V$. 
}
\label{fig:quartet_band_variational}
\end{figure}

The evolution of the lowest energy bands with increasing $t$ is illustrated in Fig.~\ref{fig:bandstructure_canonical}. The lowest band is a renormalized plaquette state. Its eigenvalue depends weakly on the wave vector, as illustrated in Fig.~\ref{fig:quartet_band_variational}. The nearly perfect agreement with a tight-binding dispersion (red dashed line)
 can be attributed to the fact that the matrix elements of the Wannier states of our basis,
$\langle SB_1{\bf n}\vert T\vert S'B_1{\bf n'}\rangle$, are only nonzero for nearest-neighbor sites ${\bf n,n'}$ (and for species $S,S'$ with different sublattice parities).
The minimum of this band is at ${\bf k=0}$. The corresponding eigenstate has $B_1$ symmetry, as the bare plaquette state, and is non-degenerate. 
Figure~\ref{fig:quartet_band_variational} shows that the energy gain due to hopping comes mostly from spreading and much less from delocalization.

\begin{figure}[t]
\centering
\includegraphics[width=\columnwidth]{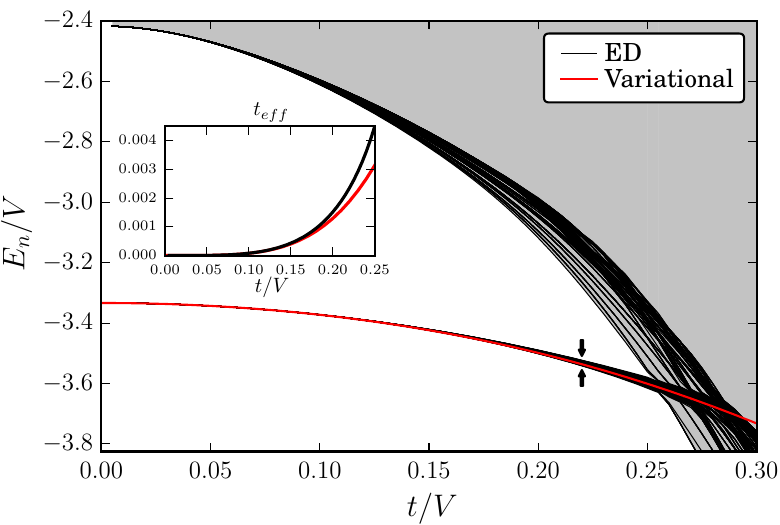}
\caption{ED spectrum for strong coupling, calculated on a $L_x=L_y=8$ lattice (black lines).
The inset shows the effective quartet hopping parameter $t_\text{eff}$ determined from the broadening of the low-energy band.
The red lines show the variational results for the center and width of the quartet band.
}
\label{fig:quartet_band_ED}
\end{figure}

ED yields translationally invariant energy eigenstates. The results shown in Fig.~\ref{fig:quartet_band_ED}, obtained for a $L_x=L_y=8$ lattice,
 confirm that a very narrow band, well separated from higher-energy states, emerges from the classical energy $E_a$ and broadens with increasing $t$. The effective hopping parameter $t_{\mbox{\scriptsize eff}}$ describing this broadening, defined as 1/8 of the measured bandwidth, increases rapidly as a function of the bare hopping parameter $t$, as shown in the inset. 
 
 The red lines in the main panel and inset of Fig.~\ref{fig:quartet_band_ED} show that the energy and width of the quartet band are well described by the variational treatment.

\subsection{Density profile}

\begin{figure}[t]
\includegraphics[width=1.02\columnwidth]{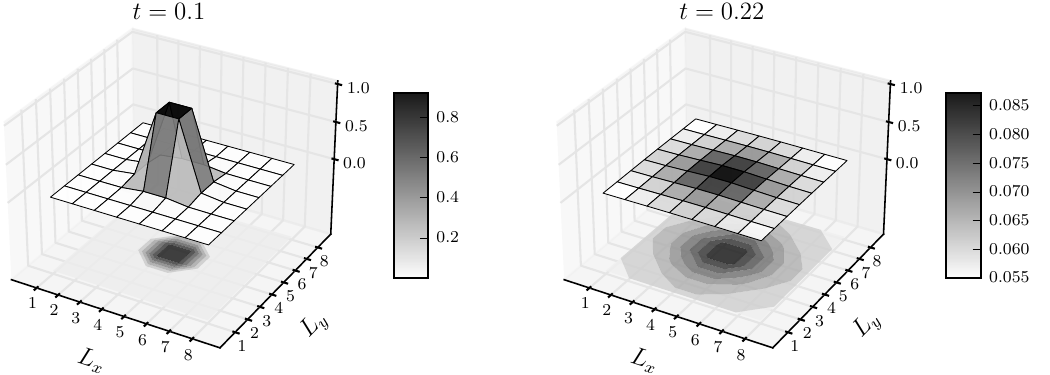}
\caption{Density profile of the ground state wave function for two hopping amplitudes in the strong-coupling regime. The left panel is for $t/V=0.1$ and the right panel for $t/V=0.22$. A small seed poential on the central plaquette is used to localize the quartet.
}
\label{fig:densities}
\end{figure}

In the translation invariant ED ground state, the quartet structure manifests itself only in the density correlation functions, but not in the density profile. To directly visualize the quartet, we add as small seed potential of  $\mathcal{O}(10^{-4})$ on four sites of a plaquette. This seed is comparable to $t_\text{eff}$ and localizes the fermions in the vicinity of the four selected sites. In Fig.~\ref{fig:densities}, we plot the resulting density profile for $t/V=0.1$ (left panel) and $t/V=0.22$ (right panel). For the smaller hopping, the wave function is clearly localized on the four sites of the plaquette, consistent with the classical picture, while for the larger hopping, the weight is spread over several sites. The true ground state without seed is a superposition of such more or less localized four-fermion states. 

\section{Transition}\label{sec:transition}

In the previous subsections, we analyzed the ground state of the four-fermion system in the weak and strong coupling limit. An interesting finding is that both the ED and variational calculations predict different ground state degeneracies and different symmetries in the two limits, namely a four-fold degenerate ground state with a decomposition $A_2\otimes B_2\otimes E$ for $V\ll t$ and a non-degenerate ground state transforming according to $B_1$ for $t\ll V$. The symmetry argument implies that these two solutions cannot be continuously connected. This result for quartets is in stark contrast to the situation for electron pairs, where the BCS-BEC crossover (e. g. in the attractive Hubbard model) describes a continuous evolution from extended to localized pairs.

\begin{figure}[t]
\includegraphics[width=\columnwidth]{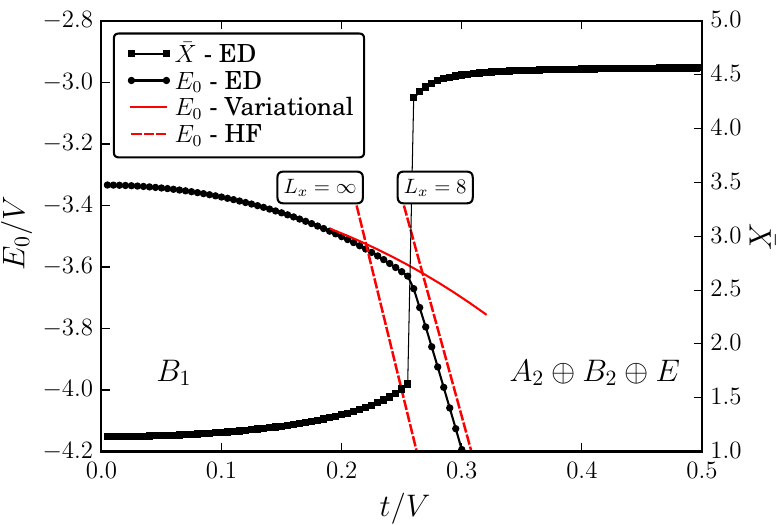}
\caption{Ground state-energy $E_0$ and spreading $\bar X$ as a function of the hopping amplitude for the coupling parameters~(\ref{eq:parameters}).
The black circles and squares show the ED results for the ground state energy and $\bar X$ on an $L_x=L_y=8$ lattice, while the red lines show Variational and Hartree-Fock results for the indicated lattice sizes.
}
\label{fig:transition}
\end{figure}

In Fig.~\ref{fig:transition} we plot the ground state energy as a function of $t/V$, together with the spreading $\bar X$ of the quartet (for $L_x=L_y=8$). This spreading is defined as the expectation value of the operator
\begin{equation}
X=\sum_{{\bf m},{\bf n}}|{\bf m}-{\bf n}|n_{\bf m}n_{\bf n}.
\end{equation} 
As can be seen in the figure, $\bar X$ exhibits a jump around $t/V=0.26$, which is the value where the ground state changes from $B_1$ to $A_2\otimes B_2\otimes E$ symmetry in the ED calculation. These results demonstrate the existence of a first order transition from a quartet bound state to a state of four unbound particles. 

We also show in Fig.~\ref{fig:transition} the ground state energies obtained from variational and Hartree-Fock calculations (red lines). These predict a transition in a similar range of $t/V$ and illustrate how the ciritical point is affected by finite-size effects. 

\section{Intermediate phase and hybrid quartet for pure attraction}
\label{sec:intermediate}

In this section we analyze the transition from the quartet ground state to the unbound four fermion state in a model with pure attraction (parametrization as in Eq.~(\ref{eq:parameters}), but with $V_3=V_4=0$).
This choice of interaction would lead to phase separation in a many-particle system, but we will show that the four-particle problem reveals the existence of a peculiar hybrid quartet ground state at intermediate coupling. A condensation of such hybrid quartets, which could be potentially realized for fine-tuned parameters, may be a possible scenario for quartet superconductivity.

\subsection{Intermediate phase}

\begin{figure}[t]
\centering
\includegraphics[width=\columnwidth]{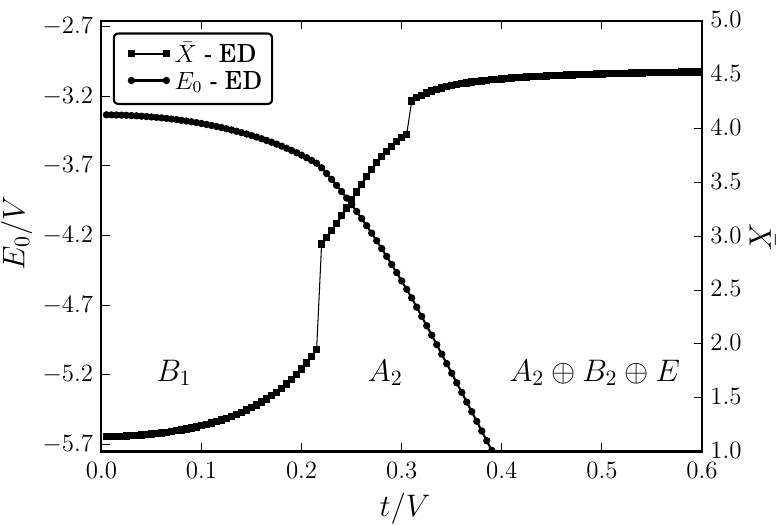}
\caption{Ground state energy $E_0$ and spreading $\bar X$ as a function of the hopping amplitude for the model with pure attraction (ED results for a $L_x=L_y=8$ lattice).
}
\label{fig:transition_attractive}
\end{figure}

In Fig.~\ref{fig:transition_attractive} we plot the ED results for the ground state energy $E_0$ and quartet spreading $\bar X$ as a function of $t/V$, for the model with pure attraction. In contrast to Fig.~\ref{fig:transition}, we now observe two jumps in $\bar X$ as the hopping parameter increases. These jumps are associated with kinks in $E_0$, indicative of level crossings. The symmetry analysis of the intermediate ground state shows that it transforms according to $A_2$, which is a different symmetry from that of the strongly bound quartet ($B_1$) and of the unbound fermions ($A_2\otimes B_2\otimes E$). These numerical results demonstrate the existence of three distinct ground states in the model with pure attraction. 

\subsection{Hybrid quartet}

\begin{figure}
\centering
\includegraphics[width=\columnwidth]{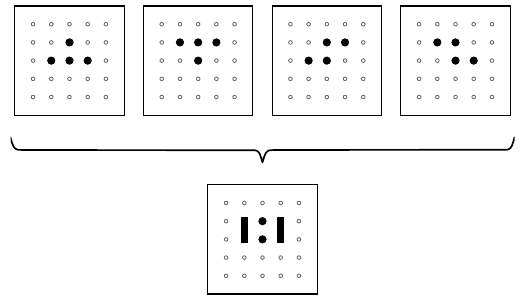}
\caption{
Relevant lattice animals in the intermediate state of the model with pure attraction, and schematic illustration of the $bc$ hybrid state (hybrid quartet).}
\label{fig:hybrid}
\end{figure}

The $\bar X$ data in Fig.~\ref{fig:transition_attractive} show that the intermediate phase corresponds to a more extended four-particle object than the tightly bound quartet, which is dominated by the plaquette state $a$ in Fig.~\ref{fig:animals}, with some admixture of $b$. On the other hand, it is more compact than the unbound four fermion state in the large hopping regime. An educated guess, which is consistent with the numerical data, is that the intermediate state is a hybrid state of lattice animals $b$ and $c$, as illustrated in Fig.~\ref{fig:hybrid}. This $bc$ hybrid with symmetry $A_2$ can be schematically represented as an extended object with two fermions in the center and two delocalized ones on the sides, as shown in the lower panel of  Fig.~\ref{fig:hybrid}.

These results can be confirmed by diagonalizing the previously mentioned $100\times 100$ Bloch matrix (Sec.~\ref{sec:zoo}). 
For small $t/V$, the ground state has $B_1$ symmetry and is dominated by the plaquette state $a$.  With increasing $t$, lattice animal $b$ acquires weight, so that one could call the ground state an $ab$ hybrid.
At some larger $t/V$, the symmetry switches from $B_1$ to $A_2$. After this transition, the weight of the $a$ character drops to zero, and the ground state can indeed be characterized as a $bc$ hybrid, since most of the weight is contributed by the $b$ and $c$ animals.

\section{Spin}
\label{sec:spin}

So far, we have limited the discussion to spinless fermions. While model \eqref{eq:ham} may be adequate for certain cold-atom systems, it may not be appropriate for electrons, especially if magnetic correlations play a role. We therefore add in this section the spin degrees of freedom and consider Hamiltonian (\ref{eq:ham}) with the hopping term
\begin{align}
T=\sum_{\langle {\bf m,n}\rangle,\sigma}\big(c_{\bf m\sigma}^\dag c_{\bf n\sigma}^{\phantom{}}+c_{\bf n\sigma}^\dag c_{\bf m\sigma}^{\phantom{}}\big)\, ,
\end{align}
where $c_{\bf n\sigma}^\dag$ ($c_{\bf n\sigma}^{\phantom{}}$) creates (annihilates) an electron with spin $\sigma=\uparrow$ or $\downarrow$ at site ${\bf n}$,
and with an additional on-site interaction term
\begin{align}\label{eq:hamiltonian}
H_{\mbox{\scriptsize int}}\rightarrow H_{\mbox{\scriptsize int}}+UD\, ,
\end{align}
where
\begin{align}
D=\sum_{\bf m}n_{\bf m\uparrow}n_{\bf m\downarrow}
\end{align}
is the number of doubly occupied sites.
The density appearing in the non-local interactions is now $n_{\bf m}:=\sum_{\sigma}n_{\bf m\sigma}$, where $n_{\bf m\sigma}=c_{\bf m\sigma}^\dag c_{\bf m\sigma}^{\phantom{}}$. The on-site coupling strength will be assumed to be non-negative, $U\ge 0$, and the other couplings $V_n$ will be chosen as in the spinless case (Eq.~\eqref{eq:parameters}). 

The addition of spin can make a qualitative difference. For example, in the weak-coupling limit there is no two-particle bound state in the spinless case, even if $\tilde{V}(0)<0$, whereas a bound state exists for two fermions with spin for infinitesimal attraction.

The addition of spin degrees of freedom increases greatly the number of linearly independent states, to 70 (instead of 1), already for four particles on the four sites of a plaquette
(16 with $D=0$, 48 with $D=1$ and 6 with $D=2$). For $t=0$ and non-local couplings as in Eq. (\ref{eq:parameters}), the plaquette states of lowest energy are those with $D=0$, even for $U=0$. They just correspond to the different configurations of four spins ($\uparrow$ or $\downarrow$) on four sites. A symmetry classification of these states is presented in Appendix \ref{app:spin}.

\subsection{Linear spreading and kinetic exchange}
In the strong-coupling limit, $V\gg t$, we could proceed as in the spinless case, using either the variational ansatz (\ref{eq:ansatz}) or a restricted zoo of lattice animals, as in Section \ref{sec:zoo}. However, in both cases, the analysis turns out to be quite involved. Therefore we limit ourselves to an adaptation of the linear ansatz (\ref{eq:linear}). 

The hopping term $T$ has now two effects. On the one hand, it again promotes spreading of the four-particle wave function. On the other hand, it leads to Anderson's kinetic exchange \cite{Anderson_59}. To deal with both effects independently, we write $T=T_1+T_2$, where $T_1$ acts only on the sites of the plaquette (${\bf n, n+i,n+i+j,n+j}$) and $T_2$ contains all the other terms, in particular those which move particles into and out of the plaquette.

We use the ansatz
\begin{align}
\vert\Psi\rangle:=\big(1+\kappa_1 T_1+\kappa_2 T_2\big)\vert\Psi_\infty\rangle\, ,
\end{align}
where $\vert\Psi_\infty\rangle$ is one of the symmetry-adapted states with singly-occupied plaquette sites, characterized by $S,S_z$ and the irreducible representations $\Gamma$ of the point group (Appendix \ref{app:spin}). 

To calculate matrix elements we notice that $T_1^2$ is directly related to the Heisenberg model, namely
\begin{align}
\langle\Psi_\infty\vert T_1^2\vert\Psi_\infty\rangle=\sum_{\langle i,j\rangle}
\langle\Psi_\infty\vert \big(1-4{\bf S}_i\cdot{\bf S}_j\big)\vert\Psi_\infty\rangle\, ,
\end{align}
where $i,j$ are plaquette sites and ${\bf S}_i$ are spin $\frac{1}{2}$ operators. The different values are, according to Eq. (\ref{eq:energies}), 
\begin{align}
\nu:=\langle\Psi_\infty\vert T_1^2\vert\Psi_\infty\rangle=
\left\{\begin{array}{rl}
0&S=2,\, \Gamma=B_1\\
8&S=1,\, \Gamma=A_2\\
4&S=1,\, \Gamma=E \,\, .\\
4&S=0,\, \Gamma=A_1\\
12&S=0,\, \Gamma=B_1
\end{array}
\right.
\end{align}
Moreover, one easily finds $\langle\Psi_\infty\vert T_2^2\vert\Psi_\infty\rangle=8$,
$\langle\Psi_\infty\vert T_1 T_2\vert\Psi_\infty\rangle=0$.
Collecting the various terms we get
\begin{align}
\langle\Psi\vert\Psi\rangle&=1+\nu\kappa_1^2+8\kappa_2^2\, ,\\
\langle\Psi\vert H_0\vert\Psi\rangle&=-\big(2\nu\kappa_1+16\kappa_2\big)t\, ,\\
\langle\Psi\vert H_{\mbox{\scriptsize int}}\vert\Psi\rangle&=E_\infty+\nu\kappa_1^2E_1+8\kappa_2^2E_2\, ,
\end{align}
where
\begin{align}
E_\infty&=E_a=4V_1+2V_2\, ,\\
E_1&=U+3V_1+2V_2\, ,\\
E_2&=E_h=2V_1+2V_2+V_3+V_4\, .
\end{align}

For $\kappa\ll 1$, $\kappa_2\ll1$ the expectation value of the Hamiltonian,
\begin{align}
\frac{\langle\Psi\vert H\vert\Psi\rangle}{\langle\Psi\vert\Psi\rangle}\approx 
E_\infty-2\nu t\kappa_1+\nu(E_1-E_\infty)\kappa_1^2-16t\kappa_2+8(E_2-E_\infty)\kappa_2^2\, ,
\end{align} 
has a minimum at
\begin{align}
\kappa_1\approx\frac{t}{E_1-E_\infty}=\frac{t}{U-V_1}\, ,\quad
\kappa_2\approx\frac{t}{E_2-E_\infty}=\frac{t}{V_3+V_4-2V_1}\, ,
\end{align}
with energy
\begin{align}
E\approx 4V_1+2V_2-\frac{\nu t^2}{U-V_1}-\frac{8t^2}{V_3+V_4-2V_1}\, .
\end{align}
Here the last term is identical to the energy gain from spreading found for spinless fermions 
(Section \ref{sec:linearspread}), while the term proportional to $\nu$ comes from kinetic exchange. The lowest energy state is a spin singlet with $B_1$ symmetry, for which $\nu=12$. We notice that an antiferromagnetic exchange is obtained even for $U=0$ (provided that $E_1<0$) and is therefore not a unique property of the repulsive Hubbard model.

\section{Conclusions}\label{sec:conclusions}

We presented a systematic investigation of (mainly spinless) fermion quartets on the square lattice. We introduced a microscopic model with short-range attractive and medium-range repulsive interactions and analyzed its four-fermion ground state using variational calculations, a Hartree-Fock type treatment and exact diagonalization. 
For the classification and symmetry analysis of the four-particle quantum states, we introduced a relevant subset of lattice animals both in real and momentum space, and used the point-group and translational symmetries.

In the weak-hopping regime the ground state is a bound state of four fermions with $B_1$ symmetry, while in the strong-hopping regime, we identified a state of four unbound fermions with a decomposition $A_2\otimes B_2\otimes E$. The different symmetries of the two ground states imply that the weak- and strong-coupling regime cannot be continuously connected, as it is the case for example in the attractive Hubbard model, where the fermion pairs show a BCS-BEC crossover. On the strong coupling side of the first order transition, the quartet bound states form a very narrow band. The kinetic energy of the quartet originates mainly from spreading, and not from delocalization, since the motion of the quartet involves a higher-order hopping process. 

While the medium range repulsion is important for preventing phase separation in a many-particle system, we can study the four-fermion problem also for a purely attractive system. In this case, both the analytical and exact diagonalization calculations revealed the existence of an intermediate phase, with yet another symmetry ($A_2$). The ground state in the intermediate phase is dominated by so-called hybrid quartet states, which can be visualized as an electron pair on neighboring sites, with a delocalized fermion on each side. The extended nature of these states should lead to an increased mobility of the hybrid quartets. 

In this study, we have not directly addressed the problem of superconductivity or quartet condensation, but the results for the single quartet allow us to draw some relevant conclusions. In particular, we have shown that in the model with canonical parameters, the bound quartets are almost localized (as evidenced by the narrow width of the quartet band). In the presence of a disorder potential, we can thus expect the quartets to easily become localized. Even without an external disorder potential, longer-ranged interactions may produce fluctuating mean fields, leading to a localization of quartets and a quartet glass. These observations provide some hints why quartet superconductivity is very rare in nature. 

With fine-tuned model parameters, it may be possible to realize a homogeneous (i.e., non-phase-separated) multi-fermion state composed of hybrid quartets. Since the hybrid quartets are more extended and presumably more mobile, the cooling of such a state could potentially lead to quartet condensation, but we leave the investigation of such a possibility as an interesting future work. More generally, the study of (multi) quartet states is interesting because the multi-fermion bound states can have a nontrivial internal structure, as exemplified by the hybrid quartet.

\appendix
\section{Construction of irreducible representations}\label{app:irreps}
\subsection{Asymmetric quartets}
Consider an asymmetric quartet state (\ref{eq:states_position}), denoted by $\vert 1\rangle$. We generate seven new states by applying $r_+$ and $\sigma_x$ with respect to some center, i.e.,
\begin{align}\label{eq:generation}
\vert 2\rangle=r_+\vert 1\rangle\, ,\quad \vert 3\rangle=r_+\vert 2\rangle\, ,\quad \vert 4\rangle=r_+\vert 3\rangle\, ,\nonumber\\
\vert 5\rangle=\sigma_x\vert 1\rangle\, ,\quad \vert 6\rangle=r_+\vert 5\rangle\, ,\quad \vert 7\rangle=r_+\vert 6\rangle\, ,\quad \vert 8\rangle=r_+\vert 7\rangle\, .
\end{align}
Any transformation of the group $D_4$ can be generated by the operations $\sigma_x$ and $r_+$, thus one obtains an 8-dimensional representation of the group, with characters 
$\chi(C_1)=8$, $\chi(C_2)=\chi(C_3)=\chi(C_4)=\chi(C_5)=0$. Equation~(\ref{eq:decomposition}) yields the decomposition (\ref{eq:decomposition1}), which corresponds to the transformation
\begin{align}\label{eq:irreps}
\vert\Gamma\rangle=\frac{1}{\sqrt{8}}\sum_{\ell=1}^8s_\ell\, \vert \ell\rangle\, ,
\end{align}
with coefficients $s_\ell$ listed in Tab.~\ref{tab:coefficients}.

\begin{table}[H]
\centering
$
\begin{array}{l|rrrrrrrr}
\Gamma&s_1&s_2&s_3&s_4&s_5&s_6&s_7&s_8\\
\hline
A_1&1&1&1&1&1&1&1&1\\
A_2&1&1&1&1&-1&-1&-1&-1\\
B_1&1&-1&1&-1&1&-1&1&-1\\
B_2&1&-1&1&-1&-1&1&-1&1\\
E&s&1&-s&-1&s&1&-s&-1\\
E'&s&1&-s&-1&-s&-1&s&1
\end{array}
$
\caption{Irreducible representations of a generic species. Here $s=\pm 1$ for the two components of the two-dimensional representations $E,E'$.}
\label{tab:coefficients}
\end{table}
\subsection{Symmetric quartets}
For quartets with a symmetry axis (such as $S=c,f,i, \dots$ in Fig.~\ref{fig:animals}) only four different states are related by symmetry, and the first line of 
Eq.~(\ref{eq:generation}) is sufficient to generate them. Proceeding as above, we obtain reducible 4-dimensional representations, which are decomposed as in Tab.~\ref{tab:irreps}.
\subsection{Non-symmorphic quartets}
The application of Eq.~(\ref{eq:generation}) to the species $b$ of Fig.~\ref{fig:animals} at each site ${\bf n}$ yields an over-complete set of states $\vert b\ell{\bf n}\rangle$. To be specific, we define
\begin{align}
\vert b1{\bf n}\rangle:=\vert {\bf n-i},{\bf n},{\bf n+i+j},{\bf n+j}\rangle\, .
\end{align}
A rotation by $\pi$ around ${\bf n}$ produces the state 
\begin{align}
\vert b3{\bf n}\rangle=\vert {\bf n+i},{\bf n},{\bf n-i-j},{\bf n-j}\rangle\, ,
\end{align}
which is equal to $\vert b1 {\bf n-j}\rangle$. Other pairs of states are mapped onto each other in the same way, and one obtains the relations
\begin{equation}
\begin{aligned}\label{eq:b-animal}
\vert b3{\bf n}\rangle=\vert b1 {\bf n-j}\rangle\, ,\\ 
\vert b4{\bf n}\rangle=\vert b2 {\bf n+i}\rangle\, ,\\ 
\vert b7{\bf n}\rangle=\vert b5 {\bf n+j}\rangle\, ,\\ 
\vert b8{\bf n}\rangle=\vert b6 {\bf n-i}\rangle\, .
\end{aligned}
\end{equation}
Therefore we can discard the states $\vert b\ell {\bf n}\rangle$ with $\ell=3,4,7,8$ and obtain a complete orthogonal set of states for the species $b$. For Bloch states, Eq.~(\ref{eq:b-animal}) leads to the relations
\begin{equation}
\begin{aligned}\label{eq:b-k}
\vert b3{\bf k}\rangle=&\, e^{-ik_y}\vert b1 {\bf k}\rangle\, ,\\ 
\vert b4{\bf k}\rangle=&\,\, e^{ik_x}\, \vert b2 {\bf k}\rangle\, ,\\ 
\vert b7{\bf k}\rangle=&\,\, e^{ik_y}\, \vert b5 {\bf k}\rangle\, ,\\ 
\vert b8{\bf k}\rangle=&\, e^{-ik_x}\vert b6 {\bf k}\rangle\, .
\end{aligned}
\end{equation}
There are four linearly independent states for each wave vector ${\bf k}$, for instance $\vert b1{\bf k}\rangle,\vert b2{\bf k}\rangle,\vert b5{\bf k}\rangle$ and $\vert b6{\bf k}\rangle$.

The Bloch states
\begin{align}
\vert S\Gamma{\bf k}\rangle=\frac{1}{\sqrt{L}}\sum_{\bf n}e^{-i{\bf k\cdot n}}\vert S\Gamma{\bf n}\rangle
\end{align}
are orthonormal for symmorphic, but not for non-symmorphic species. For instance, for $S=b$ we find
\begin{align}
\vert b\Gamma{\bf k}\rangle=\frac{1}{\sqrt{8}}\left[\left(s_1+e^{-ik_y}s_3\right)\vert b1{\bf k}\rangle
+\left(s_2+e^{ik_x}s_4\right)\vert b2{\bf k}\rangle\right.\nonumber\\
\qquad\left.+\left(s_5+e^{ik_y}s_7\right)\vert b5{\bf k}\rangle+\left(s_6+e^{-ik_x}s_8\right)\vert b6{\bf k}\rangle\right]\, ,
\end{align}
where the coefficients $s_\ell$ are listed in Tab.~\ref{tab:coefficients}. This implies
\begin{align}
\langle b\Gamma{\bf k}\vert b\Gamma{\bf k}\rangle=1\pm\frac{1}{2}\left(\cos k_x+\cos k_y\right)\, ,
\end{align}
where the upper and lower signs stand for the one- and two-dimensional irreducible representations, respectively. Moreover, the states for a given wave vector and different labels $\Gamma$ are not linearly independent, and we can limit ourselves to four of them, for instance to the four one-dimensional irreducible representations, $\Gamma=A_1,A_2,B_1,B_2$.

\section{Energy eigenvalues for weak coupling}\label{app:perturbation}
In this appendix we diagonalize the Hamiltonian in the 29-dimensional subspace illustrated in Fig.~\ref{fig:levels}. $H_0$ is diagonal with eigenvalues (\ref{eq:level1}) and (\ref{eq:level2}). 
The diagonal matrix elements of $H_{\mbox{\scriptsize int}}$ for $\vert\Psi\rangle=\vert{\bf k}_1, {\bf k}_2,{\bf k}_3,{\bf k}_4\rangle$ are
\begin{align}
L\, \langle\Psi\vert H_{\mbox{\scriptsize int}}\vert\Psi\rangle=6V({\bf 0})-\sum_{i<j}V({\bf k}_i-{\bf k}_j)\, .
\label{eq:diagonal}
\end{align}
It is convenient to introduce the notation $\widetilde{V}_i:=V({\bf q}_i)$, where ${\bf q}_0={\bf 0}$, ${\bf q}_1$ is the difference between two neighboring wave vectors, ${\bf q}_2$ that between second neighbors and so on, in close analogy to the definition of the coupling constants $V_1, ...,V_4$ on the square lattice. With this notation we can write $\sum_{i<j}V({\bf k}_i-{\bf k}_j)=\sum_i\nu_i\widetilde{V}_i$, where $\nu_1$ is the number of ``nearest-neighbor bonds'' in Fig. \ref{fig:levels}, and so on. We get
\begin{align}
L\, \langle\Psi\vert H_{\mbox{\scriptsize int}}\vert\Psi\rangle=\left\{\begin{array}{ll}6\widetilde{V}_0-3\widetilde{V}_1-2\widetilde{V}_2-\widetilde{V}_3,& \alpha\\ 
6\widetilde{V}_0-4\widetilde{V}_1-2\widetilde{V}_2,& \beta\\ 
6\widetilde{V}_0-3\widetilde{V}_1-2\widetilde{V}_2-\widetilde{V}_4,& \gamma\\ 
6\widetilde{V}_0-3\widetilde{V}_1-\widetilde{V}_2-\widetilde{V}_3-\widetilde{V}_4,& \eta\\ 
6\widetilde{V}_0-2\widetilde{V}_1-2\widetilde{V}_2-2\widetilde{V}_4,& \theta\\ 
6\widetilde{V}_0-4\widetilde{V}_2-2\widetilde{V}_3,& \phi
\end{array}
\right. .
\end{align} 

For generic coupling parameters all these expressions have different values.
This is all we need for calculating the first-order corrections to the energy for the single state $\phi$ of Fig.~\ref{fig:levels}. It is also sufficient for the four lowest-energy states 
$\alpha$, which differ in their crystal momenta, so that the off-diagonal matrix elements vanish. The same is true for the states $\beta$. This yields the energy eigenvalues
\begin{align}
E_\alpha&=E_0+\frac{1}{L}(6\widetilde{V}_0-3\widetilde{V}_1-2\widetilde{V}_2-\widetilde{V}_3)\, ,\\
E_\beta&=E_1+\frac{2}{L}(3\widetilde{V}_0-2\widetilde{V}_1-\widetilde{V}_2)\, ,\\
E_\phi&=E_1+\frac{2}{L}(3\widetilde{V}_0-2\widetilde{V}_2-\widetilde{V}_3)\, .
\end{align}

The explicit expressions for the couplings in momentum space as functions of the ``lattice constant'' $p=\frac{2\pi}{L_x}$ are
\begin{align}
\widetilde{V}_0&=4(V_1+V_2+V_3+2V_4)\, ,\nonumber\\
\widetilde{V}_1&=2[V_1+V_3+(V_1+2V_2+2V_4)\cos p+(V_3+2V_4)\cos 2p]\, ,\nonumber\\
\widetilde{V}_2&=2[V_2+2V_1\cos p+(V_2+2V_3)\cos 2p+4V_4\cos p\cos 2p]\, ,\nonumber\\
\widetilde{V}_3&=2[V_1+V_3+(V_1+2V_2+2V_4)\cos 2p+(V_3+2V_4)\cos 4p]\, ,\nonumber\\
\widetilde{V}_4&=2\left[V_4+V_1\cos p+(V_1+V_3)\cos 2p+(V_3+V_4)\cos 4p\right.\nonumber\\
&\left.+2V_2\cos 2p\cos p+2V_4\cos 4p \cos p\right]\, .
\end{align}

We turn now to the cases where off-diagonal matrix elements are involved, namely between states with two common wave vectors and with the same total momentum,
\begin{align}
L\, \langle {\bf k}_1,{\bf k}_2,{\bf k}_3,{\bf k}_4\vert H_{\mbox{\scriptsize int}}\vert {\bf k}_1,{\bf k}_2,{\bf k}_3',{\bf k}_4'\rangle=V({\bf k}_3-{\bf k}_3')-V({\bf k}_3-{\bf k}_4')\, .
\end{align}
The structure $\gamma$ of Fig.~\ref{fig:levels} represents four pairs of states with momenta $(\pm 2p,0),(0,\pm 2p)$. The pairs can be treated independently and we are left with four identical $2\times 2$ matrices with off-diagonal matrix elements
\begin{align}\label{eq:offdiagonal}
L\, \langle\Psi\vert H_{\mbox{\scriptsize int}}\vert\Psi'\rangle=\widetilde{V}_1-\widetilde{V}_3\, .
\end{align}
Therefore the originally eight-fold degenerate level is split into two four-fold degenerate levels with energies
\begin{align}
E_\gamma&=E_1+\frac{1}{L}(6\widetilde{V}_0-3\widetilde{V}_1-2\widetilde{V}_2-\widetilde{V}_4\pm\widetilde{V}_1\mp\widetilde{V}_3)\, .
\end{align}
We can proceed in the same way for the structure $\eta$, but it turns out that in this case the off-diagonal matrix elements vanish and the eight-fold degeneracy is not lifted. This is not a consequence of symmetry, therefore the degeneracy may well be lifted in higher-order perturbation theory. To first order we get
\begin{align}
E_\eta&=E_1+\frac{1}{L}(6\widetilde{V}_0-3\widetilde{V}_1-\widetilde{V}_2-\widetilde{V}_3-\widetilde{V}_4)\, .
\end{align}

Finally, we look at the four states of structure $\theta$, which all have zero momentum. It is easy to see that two ``neighboring states'' linked through a $\frac{\pi}{2}$ rotation have two common wave vectors, with matrix element (\ref{eq:offdiagonal}). On the other hand, ``opposite states'' related by a $\pi$ rotation have only the wave vector 
${\bf k}={\bf 0}$ in common, so that the matrix element between these states vanishes. This leads to a symmetric $4\times 4$ matrix $h$ with
off-diagonal matrix elements as in Eq.~(\ref{eq:offdiagonal}), except for $h_{13},\, h_{24}$, which vanish. The eigenvalues are
\begin{align}
E_\theta=E_1+\frac{2}{L}\times\left\{\begin{array}{l}3\widetilde{V}_0-\widetilde{V}_1-\widetilde{V}_2-\widetilde{V}_4,\, \mbox{twice}\\
3\widetilde{V}_0-\widetilde{V}_1-\widetilde{V}_2-\widetilde{V}_4\pm \widetilde{V}_1\mp\widetilde{V}_3\end{array}\right.  .
\end{align}

\section{Expectation values for the ansatz (\ref{eq:ansatz})}\label{app:variational}
Here we show how to evaluate the various expectation values for the ansatz (\ref{eq:ansatz}), with the plaquette state (\ref{eq:plaquette}) located at ${\bf n=0}$. In Fourier space our ansatz reads
\begin{align}\label{eq:state_momentum_rep}
\vert\Psi\rangle&=\frac{1}{L^2}\sum_{{\bf k}_1,...,{\bf k}_4}e^{-(\kappa/t)(\varepsilon_{{\bf k}_1}+...+\varepsilon_{{\bf k}_4})}e^{-i(k_{2x}+k_{3x}+k_{3y}+k_{4y})}
\, c_{{\bf k}_1}^\dag ...c_{{\bf k}_4}^\dag\vert 0\rangle\, .
\end{align}
\subsection{Norm and hopping term}\label{app:norm}

The representation (\ref{eq:state_momentum_rep}) is convenient for calculating the norm. Using Wick's theorem, we get
\begin{align}
\langle\Psi\vert\Psi\rangle=(s_0-s_2)^2[(s_0+s_2)^2-4s_1^2]\, ,
\end{align}
where
\begin{align}
s_0:=&\frac{1}{L}\sum_{\bf k}e^{-2(\kappa/t)\varepsilon_{\bf k}}\, ,\\ 
s_1:=&\frac{1}{L}\sum_{\bf k}e^{-2(\kappa/t)\varepsilon_{\bf k}}\cos k_x\, ,\\
s_2:=&\frac{1}{L}\sum_{\bf k}e^{-2(\kappa/t)\varepsilon_{\bf k}}\cos k_x\cos k_y\, .
\end{align}
For the expectation value of the hopping term we also need
\begin{align}
s_3:=&\frac{1}{L}\sum_{\bf k}e^{-2(\kappa/t)\varepsilon_{\bf k}}\cos^2 k_x\, ,\\
s_4:=&\frac{1}{L}\sum_{\bf k}e^{-2(\kappa/t)\varepsilon_{\bf k}}\cos^2 k_x\cos k_y\, .
\end{align}
We obtain
\begin{align}
\frac{\langle\Psi\vert H_0\vert\Psi\rangle}{\langle\Psi\vert\Psi\rangle}
=-8t\left\{\frac{s_1-s_4}{s_0-s_2}+\frac{(s_0+s_2)(s_1+s_4)-2s_1(s_2+s_3)}{(s_0+s_2)^2-4s_1^2}\right\}\, .
\end{align}
At first sight, it appears that we have to evaluate two-dimensional integrals to compute these expectation values. Fortunately, this is not the case, because the expressions $s_\alpha, \alpha=0,...,4$, are simple products of one-dimensional integrals, thanks to the additivity of the tight-binding spectrum, namely
\begin{align}
s_0=\tau_0^2\, ,\quad s_1=\tau_0\tau_1\, ,\quad s_2=\tau_1^2\, ,\quad s_3=\tau_0\tau_2\, ,\quad s_4=\tau_1\tau_2\, ,
\end{align}
where
\begin{align}
\tau_0:=&\frac{1}{L_x}\sum_{k_x}e^{4\kappa\cos k_x}\, ,\\ 
\tau_1:=&\frac{1}{L_x}\sum_{k_x}e^{4\kappa\cos k_x}\cos k_x\, ,\\
\tau_2:=&\frac{1}{L_x}\sum_{k_x}e^{4\kappa\cos k_x}\cos^2 k_x\, .
\end{align}
This yields the simple expression
\begin{align}\label{eq:norm}
\langle\Psi\vert\Psi\rangle=(\tau_0^2-\tau_1^2)^4
\end{align}
for the norm and Eq.~(\ref{eq:kinetic}) for the kinetic energy.

\subsection{Density}\label{app:density}
The expectation value of the density involves the ``Green functions''
\begin{align}\label{eq:Green}
g_{\bf n}=\langle 0\vert c_{\bf n}^{\phantom{}}e^{\kappa T}c_{\bf 0}^\dag\vert 0\rangle=\frac{1}{L}\sum_{\bf k}e^{-(\kappa/t)\varepsilon_{\bf k}}e^{i{\bf k\cdot n}}\, .
\end{align}
We find
\begin{align}\label{eq:density1}
\langle\Psi\vert c_{\bf n}^\dag c_{\bf n}^{\phantom{}}\vert\Psi\rangle=(\tau_0^2-\tau_1^2)^2
\Big[\tau_0^2\big(g_{\bf n}^2+g_{\bf n-i}^2+g_{\bf n-j}^2+g_{\bf n-i-j}^2\big)\nonumber\\
-2\tau_0\tau_1\big(g_{\bf n}g_{\bf n-i}+g_{\bf n}g_{\bf n-j}+g_{\bf n-i}g_{\bf n-i-j}+g_{\bf n-j}g_{\bf n-i-j}\big)\nonumber\\
+2\tau_1^2\big(g_{\bf n}g_{\bf n-i-j}+g_{\bf n-i}g_{\bf n-j}\big)\Big]\, .
\end{align}
The calculation of $g_{\bf n}$ is greatly simplified thanks to a dimensional reduction similar to that encountered in Appendix \ref{app:norm}, namely
\begin{align}\label{eq:factor}
g_{\bf n}=f_{n_x}f_{n_y}\, ,
\end{align}
where
\begin{align}
f_{n_\alpha}=\frac{1}{L_\alpha}\sum_{k_\alpha}e^{2\kappa \cos k_\alpha}\cos k_{\alpha}n_{\alpha}\, ,\quad \alpha=x,y.
\end{align}
Eqs. (\ref{eq:norm}), (\ref{eq:density1}) and (\ref{eq:factor}) yield Eq. (\ref{eq:density}).
\subsection{Interaction}\label{app:interaction}
The calculation of the interaction energy is more complicated. It involves the quantities
\begin{align}
G_{\bf n,m}^{\alpha\beta}:=g_{\bf n-a_\alpha}\, g_{\bf m-a_\beta}-g_{\bf n-a_\beta}\, g_{\bf m-a_\alpha}\, ,
\end{align}
where $g_{\bf n}$ is defined by Eq.~(\ref{eq:Green})
and the vectors ${\bf a}_\alpha,\, \alpha=1,...,4$, are the locations of the electrons in the classical reference state, i.e., 
\begin{align}
{\bf a}_1={\bf 0}\, ,\quad {\bf a}_2={\bf i}\, ,\quad {\bf a}_3={\bf i+j}\, ,\quad {\bf a}_4={\bf j}\, . 
\end{align}
The result is
\begin{widetext}
\begin{align}\label{eq:energy1}
\frac{\langle\Psi\vert H_{\mbox{\scriptsize int}}\vert\Psi\rangle}{\langle\Psi\vert\Psi\rangle}=\frac{1}{(\tau_0^2-\tau_1^2)^3}\sum_{\bf n,m}V_{\bf n,m}&\Big\{\frac{1}{2}\tau_0^2
\big[\big(G_{\bf nm}^{12}\big)^2+\big(G_{\bf nm}^{14}\big)^2+\big(G_{\bf nm}^{23}\big)^2+\big(G_{\bf nm}^{34}\big)^2\big]\nonumber\\
+\frac{1}{2}(\tau_0^2+\tau_1^2)\big[\big(G_{\bf nm}^{13}\big)^2&+\big(G_{\bf nm}^{24}\big)^2\big]+\tau_1^2\big(G_{\bf nm}^{14}G_{\bf nm}^{23}-G_{\bf nm}^{12}G_{\bf nm}^{34}\big)\nonumber\\
+\tau_0\tau_1\big[\big(G_{\bf nm}^{12}-G_{\bf nm}^{34}\big)&\big(G_{\bf nm}^{24}
-G_{\bf nm}^{13}\big)-\big(G_{\bf nm}^{14}+G_{\bf nm}^{23}\big)\big(G_{\bf nm}^{13}+G_{\bf nm}^{24}\big)\big]\Big\}\, .
\end{align}
To compute this expression, we take again advantage of the dimensional reduction (\ref{eq:factor}).

The expression (\ref{eq:energy1}) is convenient if the amplitudes $f_n$ decrease rapidly with $n$. This is the case for $\kappa\lesssim1$. For larger $\kappa$ it is preferable to transform Eq. (\ref{eq:energy1}) to momentum space, using the relation
\begin{align}
\sum_{\bf n,m}V_{\bf n,m}G_{\bf nm}^{\alpha\beta}G_{\bf nm}^{\gamma\delta}=\frac{1}{L^3}\sum_{\bf q,k,k'}V({\bf q})\, 
e^{-(\kappa/t)(\varepsilon_{\bf k}+\varepsilon_{\bf k'}+\varepsilon_{\bf k+q}+\varepsilon_{\bf k'-q})} 
f_{\alpha\beta}({\bf k,k'})f_{\gamma\delta}({\bf -k-q,-k'+q})\, ,
\end{align}
where
\begin{align}
f_{\alpha\beta}({\bf k,k'}):=e^{-i({\bf k\cdot a_\alpha}+{\bf k'\cdot a_\beta})}-e^{-i({\bf k\cdot a_\beta}+{\bf k'\cdot a_\alpha})}\, .
\end{align}
To proceed, we apply the shifts ${\bf k}\rightarrow k-\frac{\bf q}{2}$, ${\bf k'}\rightarrow k'+\frac{\bf q}{2}$ and use symmetries, such as 
${\bf q}\rightarrow {\bf -q}$, ${\bf k}\leftrightarrow {\bf k'}$. With this, we find
\begin{align}\label{eq:energy2}
\frac{\langle\Psi\vert H_{\mbox{\scriptsize int}}\vert\Psi\rangle}{\langle\Psi\vert\Psi\rangle}=\frac{2}{(\tau_0^2-\tau_1^2)^3}
\frac{1}{L}\sum_{\bf q}V({\bf q})\left\{-\tau_0^2\big[v^2(q_x)u^2(q_y)+u^2(q_x)v^2(q_y)\big]\right.\nonumber\\
-(\tau_0^2+3\tau_1^2)v^2(q_x)v^2(q_y)+u^2(q_x)\big[\tau_0^2u^2(q_y)+\tau_1^2v^2(q_y)\big]\cos q_x\nonumber\\
+u^2(q_y)\big[\tau_0^2u^2(q_x)+\tau_1^2v^2(q_x)\big]\cos q_y+(\tau_0^2+\tau_1^2)u^2(q_x)u^2(q_y)\cos q_x\cos q_y\nonumber\\
+4\tau_0\tau_1v(q_x)v(q_y)\big[u(q_x)v(q_y)\cos\frac{q_x}{2}+v(q_x)u(q_y)\cos\frac{q_y}{2}\big]\nonumber\\
\left.-4\tau_0\tau_1u(q_x)u(q_y)\big[u(q_x)v(q_y)\cos q_x\cos\frac{q_y}{2}+v(q_x)u(q_y)\cos\frac{q_x}{2}\cos q_y\big]\right\}\, ,
\end{align} 
\end{widetext}
 where ($\alpha=x,y$)
 \begin{align}\label{eq:uv}
 &u(q_\alpha):=\frac{1}{L_\alpha}\sum_{k_\alpha}e^{4\kappa\cos\frac{q_\alpha}{2}\cos k_\alpha}\, ,\nonumber\\
 &v(q_\alpha):=\frac{1}{L_\alpha}\sum_{k_\alpha}e^{4\kappa\cos\frac{q_\alpha}{2}\cos k_\alpha}\cos k_\alpha\, .
 \end{align}
 For periodic boundary conditions, the wave vectors ${\bf q}$ are given by
 \begin{align}
 {\bf q}=\frac{2\pi}{L_x}(\nu_x,\nu_y)
 \end{align}
with integers $\nu_\alpha$ in the range $-\frac{L_x}{2}<\nu_\alpha\le\frac{L_x}{2}$. Due to the momentum shifts by $\pm\frac{\bf q}{2}$, the wave vectors $k_\alpha$ in the sums (\ref{eq:uv}) are ``even'' or ``odd'', depending on ${\bf q}$,
\begin{align}
k_\alpha=\frac{\pi}{L_x}\left\{\begin{array}{ll}2\mu_\alpha,&\mbox{ if }\nu_\alpha\mbox{ is even,}\\
(2\mu_\alpha+1),&\mbox{ if }\nu_\alpha\mbox{ is odd.}\end{array}\right.
\end{align}
The integers $\mu_\alpha$ can again be chosen in the range $-\frac{L_x}{2}<\mu_\alpha\le\frac{L_x}{2}$ because all terms in Eq. (\ref{eq:energy2}) are $2\pi$-periodic.

\section{Spin states on the plaquette}\label{app:spin}
Here we classify the 16 different spin configurations on the plaquette in terms of the total spin ($S,S_z$) and the irreducible representations $\Gamma$ of the point group, using the notation
\begin{align}
\vert\sigma_1\sigma_2\sigma_3\sigma_4\rangle:=
c_{{\bf n}\sigma_1}^\dag c_{{\bf n+i}\sigma_2}^\dag c_{{\bf n+i+j}\sigma_3}^\dag
 c_{{\bf n+j}\sigma_4}^\dag\vert 0\rangle\, .
\end{align}
There are 5 states with $S=2$, 9 with $S=1$ and 2 with $S=0$. Those with $S_z=S$, written as
$\vert S\Gamma\rangle$, are
\begin{align}\label{eq:spinstates}
\vert 2B_1\rangle&=\vert\!\!\uparrow\uparrow\uparrow\uparrow\rangle\, ,\nonumber\\
\vert 1A_2\rangle&=\frac{1}{2}\big(\vert\!\!\downarrow\uparrow\uparrow\uparrow\rangle-\vert\!\!\uparrow\downarrow\uparrow\uparrow\rangle
+\vert\!\!\uparrow\uparrow\downarrow\uparrow\rangle-\vert\!\!\uparrow\uparrow\uparrow\downarrow\rangle\big)\, ,\nonumber\\
\vert 1E\rangle_\pm&=\frac{1}{2}\big(\vert\!\!\downarrow\uparrow\uparrow\uparrow\rangle\mp\vert\!\!\uparrow\downarrow\uparrow\uparrow\rangle
-\vert\!\!\uparrow\uparrow\downarrow\uparrow\rangle\pm\vert\!\!\uparrow\uparrow\uparrow\downarrow\rangle\big)\, ,\nonumber\\
\vert 0A_1\rangle&=\frac{1}{2}\big(\vert\!\!\downarrow\downarrow\uparrow\uparrow\rangle-\vert\!\!\downarrow\uparrow\uparrow\downarrow\rangle
+\vert\!\!\uparrow\uparrow\downarrow\downarrow\rangle-\vert\!\!\uparrow\downarrow\downarrow\uparrow\rangle\big)\, ,\nonumber\\
\vert 0B_1\rangle&=\frac{1}{2\sqrt{3}}\big(2\vert\!\!\uparrow\downarrow\uparrow\downarrow\rangle+2\vert\!\!\downarrow\uparrow\downarrow\uparrow\rangle
-\vert\!\!\downarrow\downarrow\uparrow\uparrow\rangle-\vert\!\!\downarrow\uparrow\uparrow\downarrow\rangle
-\vert\!\!\uparrow\uparrow\downarrow\downarrow\rangle-\vert\!\!\uparrow\downarrow\downarrow\uparrow\rangle\big)\, ,
\end{align}
where the subscript $\pm$ distinguishes the two $E$-states according to the eigenvalues $\pm 1$ of the reflection operator $\sigma_x$.

The states (\ref{eq:spinstates}) are eigenstates of the Heisenberg Hamiltonian
\begin{align}
H=J\sum_{\langle i,j\rangle}\Big({\bf S}_i\cdot {\bf S}_j -\frac{1}{4}\Big)
\end{align}
with eigenvalues
\begin{align}\label{eq:energies}
E=\left\{\begin{array}{cl} 
  0&S=2,\, \Gamma=B_1\\
-2J&S=1,\, \Gamma=A_2\\
-J&S=1,\, \Gamma=E\\
-J&S=0,\, \Gamma=A_1\\
-3J&S=0,\, \Gamma=B_1
\end{array}\right. .
\end{align}
The ground state is a singlet, and transforms according to the irreducible representation $B_1$.

\bibliographystyle{apsrev}
\bibliography{quartet}

\end{document}